\documentclass[a4paper,10pt]{revtex4}
\usepackage{graphicx}  
\usepackage{amsmath}   
\usepackage{amssymb}   
\usepackage{bm} 
\usepackage{dcolumn}
\usepackage{color}
\usepackage{mathrsfs}
\usepackage{amsfonts}
\usepackage{varioref}
\usepackage{mathrsfs}
\usepackage{graphicx}
\usepackage{latexsym}
\usepackage{amsmath}
\usepackage{amssymb}
\usepackage{textcomp}
\usepackage{amsbsy}
\usepackage{graphics}
\usepackage{epstopdf}
\usepackage{color}
\usepackage[caption=false]{subfig}

\RequirePackage[colorlinks,citecolor=blue,urlcolor=magenta,linkcolor=blue]{hyperref}
\input epsf

\allowdisplaybreaks[4]

\begin{document}

\tolerance=5000

\title{A non-singular generalized entropy and its implications on bounce cosmology}

\author{Sergei~D.~Odintsov$^{1,2}$\,\thanks{odintsov@ieec.uab.es},
Tanmoy~Paul$^{3}$\,\thanks{pul.tnmy9@gmail.com}} \affiliation{
$^{1)}$ ICREA, Passeig Luis Companys, 23, 08010 Barcelona, Spain\\
$^{2)}$ Institute of Space Sciences (ICE, CSIC) C. Can Magrans s/n, 08193 Barcelona, Spain\\
$^{3)}$ Department of Physics, Chandernagore College, Hooghly - 712 136, India.}


\tolerance=5000

\begin{abstract}
We propose a new five-parameter entropy function that proves to be singular-free during the entire cosmic evolution of the universe, 
and at the same time, also generalizes the Tsallis, Barrow, R\'{e}nyi, Sharma-Mittal, 
Kaniadakis and Loop Quantum Gravity entropies for suitable limits of the parameters. In particular, all 
the above mentioned known entropies become singular (or diverge) 
when the Hubble parameter vanishes in course of the universe's evolution (for instance, in bounce cosmology at the instant of bounce), while 
the newly proposed entropy function seems to be singular-free even at $H = 0$ (where $H$ represents the Hubble parameter of the universe). 
Such $non-singular$ behaviour of the entropy function becomes useful in describing bouncing scenario, in which case, the universe undergoes through 
$H = 0$ at the instant of bounce. It turns out that the entropic cosmology corresponding to the singular-free generalized entropy naturally allows 
symmetric bounce scenarios, such as -- exponential bounce and quasi-matter bounce scenario respectively. In the case of exponential bounce, the perturbation 
modes are in the super-Hubble domain at the distant past, while for the quasi-matter bounce, the perturbation modes generate in the deep sub-Hubble regime 
far before the bounce and hence resolves the ``horizon issue''. Based on this fact, we perform a detailed perturbation analysis for 
the quasi-matter bounce in the present context of singular-free entropic cosmology. As a result, the primordial observable 
quantities like the spectral tilt for the curvature perturbation ($n_s$) and the tensor-to-scalar ratio ($r$) are found to depend on the entropic 
parameters, as expected. The theoretical expectations of $n_s$ and $r$ turn out to be simultaneously compatible with the 
recent Planck data for suitable ranges of the entropic parameters, which in turn ensures the viability of the entropic bounce scenario. 
Furthermore the entropic cosmology in the present context is shown to be equivalent with the generalized holographic cosmology where the holographic 
cut-offs are determined in terms of either future horizon and its derivative or the particle horizon and its derivative.
\end{abstract}

\maketitle

\section{Introduction}

Entropy, one of the important and fundamental quantities in physics, seems to depend on the physical system under consideration. For example, 
the entropy in classical thermodynamics is found to be proportional to the volume of the system, while the black hole entropy 
is proportional to the horizon area. This may indicate that we do not possibly understand the fundamental construction of entropy till now, or, 
there possibly exists a generalized form of entropy that is true irrespective of the choices of the system(s). 

The black body radiation of a black hole is regarded as one of the remarkable discoveries of theoretical physics, which is described by a 
finite temperature and by a Bekenstein-Hawking entropy function \cite{Bekenstein:1973ur,Hawking:1975vcx} (see 
\cite{Bardeen:1973gs,Wald:1999vt} for extensive reviews). The distinctive property of the Bekenstein-Hawking entropy is that the entropy function 
is proportional to the horizon area of the black hole, unlike to the classical thermodynamics where the entropy is directly proportional ton the volume 
of the system under consideration. Such unusual behaviour of the black hole entropy results to the recently proposed various forms of entropy 
functions other than the Bekenstein-Hawking one. For instance, the Tsallis \cite{Tsallis:1987eu} and the R\'{e}nyi \cite{Renyi} 
entropies have been proposed depending on the 
non-additive statistics of the system. Recently Barrow proposed an entropy function in \cite{Barrow:2020tzx}, 
which encodes the fractal structure of a black hole that may originate from quantum grvaity effects. 
Beside these, the Sharma-Mittal entropy \cite{SayahianJahromi:2018irq}, the Kaniadakis entropy \cite{Kaniadakis:2005zk,Drepanou:2021jiv} and 
the Loop Quantum Gravity entropy \cite{Majhi:2017zao,Liu:2021dvj} are some well known entropies proposed so far. All of these 
entropies reduce to the Bekenstein-Hawking entropy at certain limit, and moreover, they are monotonic increasing function with respect to the 
Bekenstein-Hawking entropy variable. 

The thermodynamics laws with Bekenstein-Hawking entropy are extended to the cosmology sector, by which, one can achieve the usual Friedmann equations.
In accordance, we may argue that the cosmological field equations have a thermodynamic nature. In this regard, 
the holographic cosmology, initiated in \cite{Witten:1998qj,Susskind:1998dq,Fischler:1998st}, 
gained a lot of attention as it is intimately connected to entropy construction. 
The entropic cosmology corresponding to different entropy functions (for example the Tsallis entropic cosmology, the R\'{e}nyi entropic cosmology etc.) 
are proved to be equivalent with generalized holographic cosmology where the respective cut-offs are determined in terms of either particle horizon and 
its derivative or the future horizon and its derivative \cite{Nojiri:2021iko,Nojiri:2021jxf}. 
The holographic cosmology stands to be one of the useful theories 
in describing the dark energy era of the universe, in particular, the dark energy density in a holographic dark energy (HDE) model is sourced from holographic 
principle, rather than put by hand in the Lagrangian 
\cite{Li:2004rb,Li:2011sd,Wang:2016och,Pavon:2005yx,Nojiri:2005pu,Landim:2022jgr,Zhang:2005yz,Guberina:2005fb,Elizalde:2005ju,
Ito:2004qi,Gong:2004cb,Nojiri:2022nmu,BouhmadiLopez:2011xi,Malekjani:2012bw,Khurshudyan:2016gmb,Landim:2015hqa,
Gao:2007ep,Li:2008zq,Anagnostopoulos:2020ctz,Zhang:2005hs,Li:2009bn,Feng:2007wn,Zhang:2009un,Lu:2009iv,
Micheletti:2009jy,Mukherjee:2017oom,Nojiri:2017opc,Nojiri:2019skr,
Saridakis:2020zol,Barrow:2020kug,Adhikary:2021xym,Srivastava:2020cyk,Bhardwaj:2021chg,Chakraborty:2020jsq,Sarkar:2021izd}. Beside the HDE models, the holographic cosmology is extended to explain 
the early inflationary phase of the universe where the primordial quantities turn out to be compatible with the recent observational data 
\cite{Nojiri:2022aof,Nojiri:2022dkr,Horvat:2011wr,Nojiri:2019kkp,Paul:2019hys,Bargach:2019pst,Elizalde:2019jmh,Oliveros:2019rnq,Mohammadi:2022vru,Chakraborty:2020tge}, 
and more 
importantly, the holographic cosmology shows to be useful to unify the early inflation with the dark energy era of the universe 
\cite{Nojiri:2022dkr,Nojiri:2020wmh}. From a different 
theoretical construction, the holographic energy density and the corresponding pressure help to violate the null energy condition during the early phase of the 
universe, which in turn provides a non-singular bouncing scenario from a holographic point of view \cite{Nojiri:2019yzg,Brevik:2019mah}. 
All of such literatures reveal the immense interest on holographic as well as on entropic cosmology corresponding to different entropy constructions. 

Based on the above arguments, the question that naturally arises is following:
\begin{itemize}
\item Does there exist a generalized entropy function that generalizes the known entropy functions proposed so far, like the 
Tsallis, R\'{e}nyi, Barrow, Sharma-Mittal, Kaniadakis and Loop Quantum Gravity entropies?
\end{itemize}
Possible explanations of this question has been given in \cite{Nojiri:2022aof,Nojiri:2022dkr,Nojiri:2022ljp}. 
In particular, some of our authors proposed a 6-parameter entropy function 
that generalizes all the aforementioned known entropies \cite{Nojiri:2022aof}. However in \cite{Nojiri:2022dkr}, 
we proposed a different form of generalized entropy 
containing less parameters, particularly having four-parameters, which is also able to generalize the known entropies from the Tsallis to Loop Quantum 
Gravity entropies. In this regard, we give the following conjecture: 
``The minimum number of parameters required in a generalized entropy function that can generalize 
all the aforementioned entropies is equal to four''. Moreover, the generalized entropy with four parameters proved to be useful to explain 
the early inflation and the dark energy era of the universe in an unified manner. However at this stage, it deserves mentioning that 
all of these generalized entropies mentioned in \cite{Nojiri:2022aof,Nojiri:2022dkr} 
become singular (or diverge) during the cosmic evolution of the universe, particularly 
when the Hubble parameter of the universe vanishes (for instance, 
in a bounce cosmology -- when the Hubble parameter vanishes at the instant of the bounce). Such diverging behaviour also shows 
to all the known entropies, due to the fact that the Bekenstein-Hawking entropy (given by $S = \frac{\pi}{GH^2}$) itself diverges at 
$H = 0$. Thus an immediate question is given by:
\begin{itemize}
\item Does there exist a singular-free generalized entropy that generalizes the 
Tsallis, R\'{e}nyi, Barrow, Sharma-Mittal, Kaniadakis and Loop Quantum Gravity entropies, and at the same time, is also singular-free 
during the cosmological evolution of the universe (even at $H = 0$ where $H$ represents the Hubble parameter) ?
\item If so, then what are the cosmological implications of such non-singular entropy function ?
\end{itemize}
We will try to address these questions in the present work. Here we would like to mention that if such non-singular entropy exists, then it 
proves to be useful in describing the bounce scenario where the universe undergoes through $H = 0$ at the instant of the bounce, unlike to all the 
known entropies (including the generalized entropies proposed in \cite{Nojiri:2022aof,Nojiri:2022dkr,Nojiri:2022ljp}) 
which diverge at $H = 0$ and hence they are unacceptable 
in the context of bounce cosmology. Based on this argument, we will address the possible implications of a non-singular generalized entropy 
to bounce cosmology.

\section{Search for a singuar-free generalized entropy}\label{sec-gen-entropy}

In this section, we will propose a generalized entropy function which is singular-free and 
can lead to various known entropy functions proposed so far. Here it deserves mentioning that in one of our 
previous works \cite{Nojiri:2022dkr}, we have proposed a generalized four parameter entropy function given by,
\begin{eqnarray}
 S_\mathrm{g}^{(s)}\left[\alpha_+,\alpha_-,\beta,\gamma \right] = \frac{1}{\gamma}\left[\left(1 + \frac{\alpha_+}{\beta}~S\right)^{\beta} 
 - \left(1 + \frac{\alpha_-}{\beta}~S\right)^{-\beta}\right] \,,
\label{gen-entropy-earlier}
\end{eqnarray}
that converges to the known entropies (in particular, the Bekenstein-Hawking entropy, Tsallis entropy, Barrow entropy, R\'{e}nyi entropy, 
Kaniadakis entropy, Sharma-Mittal entropy and the entropy in the context of Loop Quantum gravity) for suitable choices of the parameters. 
Moreover the entropic cosmology corresponds to the $S_\mathrm{g}^{(s)}$ results to a viable unification of inflation to the dark energy epoch which 
are well consistent with the observational constraints, see \cite{Nojiri:2022dkr}. Despite these successes, 
the above entropy function $S_\mathrm{g}^{(s)}$ seems to be plagued with singularity for certain cosmological evolution of the universe, 
in particular, in the context of bounce cosmology. The demonstration goes as follows: 
in the right hand side of Eq.(\ref{gen-entropy-earlier}), $S = A/\left(4G\right)$ represents the Bekenstein-Hawking entropy, 
where $A = 4\pi r_h^2$ is the area of the horizon and $r_h$ is the horizon radius. Using $r_h = 1/H$, with $H$ being the Hubble parameter 
of the universe, one can represent the Bekenstein-Hawking entropy as $S = \pi/\left(GH^2\right)$. In effect, the $S_\mathrm{g}^{(s)}$ from 
Eq.(\ref{gen-entropy-earlier}) is equivalently written as,
\begin{eqnarray}
 S_\mathrm{g}^{(s)}\left[\alpha_+,\alpha_-,\beta,\gamma \right] = \frac{1}{\gamma}\left[\left(1 + \frac{\pi\alpha_+}{\beta GH^2}\right)^{\beta} 
 - \left(1 + \frac{\pi\alpha_-}{\beta GH^2}\right)^{-\beta}\right] \,.
\label{gen-entropy-earlier-1}
\end{eqnarray}
Clearly in the context of bounce cosmology, $S_\mathrm{g}^{(s)}$ becomes singular (or diverges) at the instant of bounce when 
the Hubble parameter of the universe vanishes. Therefore in a bounce scenario, the generalized entropy function shown 
in Eq.(\ref{gen-entropy-earlier}) is not physical, and thus, we need to search for a generalized entropy function which can lead 
to various known entropy functions and is non-singular for the entire cosmological evolution of the universe.

We propose a new singular-free entropy function given by,
\begin{eqnarray}
S_\mathrm{g}\left[\alpha_{\pm},\beta,\gamma,\epsilon\right] = \frac{1}{\gamma}\bigg[\left\{1 + \frac{1}{\epsilon}\tanh\left(\frac{\epsilon \alpha_+}{\beta}~S\right)\right\}^{\beta} 
- \left\{1 + \frac{1}{\epsilon}\tanh\left(\frac{\epsilon \alpha_-}{\beta}~S\right)\right\}^{-\beta}\bigg]~~,
\label{gen-entropy}
\end{eqnarray}
where $\alpha_{\pm}$, $\beta$, $\gamma$ and $\epsilon$ are the parameters which are considered to be positive, and $S$ symbolizes 
the Bekenstein-Hawking entropy. In regard to the number of parameters, we propose a conjecture at the end of this section. 
First we demonstrate that the above entropy function remains finite, and thus is non-singular, during the 
whole cosmological evolution of a bouncing universe. Due to $S = \pi/\left(GH^2\right)$, the $S_\mathrm{g}$ from Eq.(\ref{gen-entropy}) 
can be re-written as,
\begin{eqnarray}
S_\mathrm{g}\left[\alpha_{\pm},\beta,\gamma,\epsilon\right] = 
\frac{1}{\gamma}\bigg[\left\{1 + \frac{1}{\epsilon}\tanh\left(\frac{\epsilon \pi\alpha_+}{\beta GH^2}\right)\right\}^{\beta} 
- \left\{1 + \frac{1}{\epsilon}\tanh\left(\frac{\epsilon \pi\alpha_-}{\beta GH^2}\right)\right\}^{-\beta}\bigg]~~,
\label{gen-entropy-1}
\end{eqnarray}
which, due to the presence of $\tanh\left(\frac{\epsilon \pi\alpha_{\pm}}{\beta GH^2}\right)$, is clearly finite at $H = 0$. In particular, 
the $S_\mathrm{g}$ takes the following form at the instant of bounce:
\begin{eqnarray}
S_\mathrm{g}\left[\alpha_{\pm},\beta,\gamma,\epsilon\right] = 
\frac{1}{\gamma}\bigg[\left\{1 + \frac{1}{\epsilon}\right\}^{\beta} 
- \left\{1 + \frac{1}{\epsilon}\right\}^{-\beta}\bigg]~~.
\label{gen-entropy-2}
\end{eqnarray}
Having demonstrated the non-singular behaviour of the entropy function, we now show that $S_\mathrm{g}$ of Eq.(\ref{gen-entropy}), 
for suitable choices of the parameters, 
reduces to various known entropies-- i.e the Bekenstein-Hawking entropy, Tsallis entropy, Barrow entropy, R\'{e}nyi entropy, 
Kaniadakis entropy, Sharma-Mittal entropy and the entropy in the context of Loop Quantum gravity. Such entropies are constructed from different viewpoints. For instance, the Tsallis entropy is given by $S_T = S^{\delta}$ (with $\delta$ being the Tsallis exponent and $S$ represents the Bekenstein-Hawking entropy) which is useful for the systems having long range interactions where the Boltzmann-Gibbs entropy is not applied \cite{Tsallis:1987eu}. Clearly for $\delta = 1$, the Tsallis entropy converges to the Bekenstein-Hawking limit, however for $\delta \neq 1$, it lacks of additivity. The Barrow entropy has almost the same form as the Tsallis one, however the motivation for introducing the Barrow entropy is different. In particular, the Barrow entropy is given by $S_\mathrm{B} = \left(\frac{A}{A_\mathrm{Pl}} \right)^{1+\Delta/2}$ where $A$ is the usual black hole horizon area, $A_\mathrm{Pl} = 4G$ is the Planck area and $\Delta$ is the Barrow exponent which measures the fractal features on the black hole structure that may generate from quantum gravitational effects \cite{Barrow:2020tzx}. On other side, the R\'{e}nyi entropy is of the form of $S_\mathrm{R} = \frac{1}{\alpha} \ln \left( 1 + \alpha S \right)$ which has the Bekenstein-Hawking limit for $\alpha \rightarrow 0$ \cite{Renyi}. The R\'{e}nyi entropy was proposed as an index specifying the amount of information. The Sharma-Mittal entropy has been introduced as a possible combination of Tsallis and R\'{e}nyi entropies \cite{SayahianJahromi:2018irq}, and is given by $S_{SM} = \frac{1}{R}\left[\left(1 + \delta ~S\right)^{R/\delta} - 1\right]$. The Kaniadakis entropy has the form : $S_K = \frac{1}{K}\sinh{\left(KS\right)}$ which can be regarded as a possible generalization of the Boltzmann-Gibbs entropy arising in relativistic statistical systems \cite{Kaniadakis:2005zk,Drepanou:2021jiv}. In the context of Loop Quantum Gravity, one may get the following entropy as,
$S_q = \frac{1}{\left(1-q\right)}\left[\mathrm{e}^{(1-q)\Lambda(\gamma_0)S} - 1\right]$ where $\gamma_0$ is known as the Barbero-Immirzi parameter and $q$ is the entropic index that quantifies the probability of frequent events \cite{Majhi:2017zao,Liu:2021dvj}.

\begin{itemize}
\item For $\epsilon \rightarrow 0$, $S_\mathrm{g}$ tends to the following form,
\begin{eqnarray}
 S_\mathrm{g} = \frac{1}{\gamma}\left[\left(1 + \frac{\alpha_+}{\beta}~S\right)^{\beta} 
 - \left(1 + \frac{\alpha_-}{\beta}~S\right)^{-\beta}\right]~~,
\end{eqnarray}
which, for $\alpha_+ \rightarrow \infty$ and $\alpha_- = 0$, becomes
\begin{align}
S_\mathrm{g} = \frac{1}{\gamma}\left(\frac{\alpha_+}{\beta}\right)^{\beta}S^{\beta} \,.\nonumber
\end{align}
Further considering $\gamma = \left(\alpha_+/\beta\right)^{\beta}$, the generalized entropy $S_\mathrm{g}$ reduces to
\begin{align}
 S_\mathrm{g} = S^{\beta} \,.\label{Tsallis-Barrow}
\end{align}
This resembles to the Tsallis entropy \cite{Tsallis:1987eu} or to the Barrow entropy \cite{Barrow:2020tzx} 
with $\beta = \delta$ or $\beta = 1 + \Delta$ respectively.

\item The limit $\epsilon \rightarrow 0$ results to the following form of $S_\mathrm{g}$ as,
\begin{eqnarray}
 S_\mathrm{g} = \frac{1}{\gamma}\left[\left(1 + \frac{\alpha_+}{\beta}~S\right)^{\beta} 
 - \left(1 + \frac{\alpha_-}{\beta}~S\right)^{-\beta}\right]~~,
\end{eqnarray}
which, for $\epsilon \rightarrow 0$, $\alpha_- = 0$, $\beta \rightarrow 0$ and $\frac{\alpha_+}{\beta} \rightarrow \mathrm{finite}$, tends to,
\begin{align}
 S_\mathrm{g} = \frac{1}{\gamma}\left[\left(1 + \frac{\alpha_+}{\beta}~S\right)^{\beta} - 1\right] 
 = \frac{1}{\gamma}\left[\exp{\left\{\beta\ln{\left(1 + \frac{\alpha_+}{\beta}~S\right)}\right\}} - 1\right] 
 \approx \frac{1}{\left(\gamma/\beta\right)} \ln{\left(1 + \frac{\alpha_+}{\beta}~S\right)} \,.\label{Renyi}
\end{align}
This has the form of R\'{e}nyi entropy \cite{Renyi} with the identifications 
$\gamma = \alpha_+$ and $\frac{\alpha_+}{\beta} = \alpha$, in particular,
\begin{align}
 S_\mathrm{g} = \frac{1}{\alpha}\ln{\left(1 + \alpha~S\right)} \,.
\end{align}

\item For $\epsilon \rightarrow 0$ and $\alpha_- \rightarrow 0$, the non-singular generalized entropy converges to the following form,
\begin{eqnarray}
 S_\mathrm{g}&=&
 \lim_{\epsilon\rightarrow 0} \frac{1}{\gamma}\bigg[\left\{1 + \frac{1}{\epsilon}\tanh\left(\frac{\epsilon \alpha_+}{\beta}~S\right)\right\}^{\beta} 
- \left\{1 + \frac{1}{\epsilon}\tanh\left(\frac{\epsilon \alpha_-}{\beta}~S\right)\right\}^{-\beta}\bigg] 
= \frac{1}{\gamma}\left[\left(1 + \frac{\alpha_+}{\beta}~S\right)^{\beta} 
 - \left(1 + \frac{\alpha_-}{\beta}~S\right)^{-\beta}\right]\nonumber\\
 &\longrightarrow&\lim_{\alpha_- \rightarrow 0} \frac{1}{\gamma}\left[\left(1 + \frac{\alpha_+}{\beta}~S\right)^{\beta} 
 - \left(1 + \frac{\alpha_-}{\beta}~S\right)^{-\beta}\right] 
 = \frac{1}{\gamma}\left[\left(1 + \frac{\alpha_+}{\beta}~S\right)^{\beta} - 1\right]\label{SM}
\end{eqnarray}
Therefore with $\gamma = R$, $\alpha_+ = R$ and $\beta = R/\delta$, the above form of $S_\mathrm{g}$ becomes similar to the 
Sharma-Mittal entropy \cite{SayahianJahromi:2018irq}.

\item For $\epsilon \rightarrow 0$, 
$\beta \rightarrow \infty$, $\alpha_+ = \alpha_- = \frac{\gamma}{2} = K$, one may write Eq.~(\ref{gen-entropy}) as,
\begin{eqnarray}
S_\mathrm{g}&=&
\lim_{\epsilon\rightarrow 0} \frac{1}{\gamma}\bigg[\left\{1 + \frac{1}{\epsilon}\tanh\left(\frac{\epsilon \alpha_+}{\beta}~S\right)\right\}^{\beta} 
- \left\{1 + \frac{1}{\epsilon}\tanh\left(\frac{\epsilon \alpha_-}{\beta}~S\right)\right\}^{-\beta}\bigg] 
= \frac{1}{\gamma}\left[\left(1 + \frac{\alpha_+}{\beta}~S\right)^{\beta} 
 - \left(1 + \frac{\alpha_-}{\beta}~S\right)^{-\beta}\right]\nonumber\\
&\longrightarrow&\frac{1}{2K}\lim_{\beta \rightarrow \infty}\left[\left(1 + \frac{K}{\beta}~S\right)^{\beta} 
 - \left(1 + \frac{K}{\beta}~S\right)^{-\beta}\right] 
 = \frac{1}{2K}\left[\mathrm{e}^{KS} - \mathrm{e}^{-KS}\right] = \frac{1}{K}\sinh{\left(KS\right)}
\label{Kaniadakis}
\end{eqnarray}
that is similar to the Kaniadakis entropy \cite{Kaniadakis:2005zk,Drepanou:2021jiv}.

\item Finally, with $\epsilon \rightarrow 0$, $\alpha_- \rightarrow 0$, $\beta \rightarrow \infty$ 
and $\gamma = \alpha_+ = (1-q)$, Eq.~(\ref{gen-entropy}) yields to,
\begin{eqnarray}
 S_\mathrm{g}&=&
 \lim_{\epsilon\rightarrow 0} \frac{1}{\gamma}\bigg[\left\{1 + \frac{1}{\epsilon}\tanh\left(\frac{\epsilon \alpha_+}{\beta}~S\right)\right\}^{\beta} 
- \left\{1 + \frac{1}{\epsilon}\tanh\left(\frac{\epsilon \alpha_-}{\beta}~S\right)\right\}^{-\beta}\bigg] 
= \frac{1}{\gamma}\left[\left(1 + \frac{\alpha_+}{\beta}~S\right)^{\beta} 
 - \left(1 + \frac{\alpha_-}{\beta}~S\right)^{-\beta}\right]\nonumber\\
 &\longrightarrow&\lim_{\alpha_- \rightarrow 0} \frac{1}{\gamma}\left[\left(1 + \frac{\alpha_+}{\beta}~S\right)^{\beta} 
 - \left(1 + \frac{\alpha_-}{\beta}~S\right)^{-\beta}\right] 
 = \frac{1}{\gamma}\left[\left(1 + \frac{\alpha_+}{\beta}~S\right)^{\beta} - 1\right]\nonumber\\
 &\longrightarrow&\frac{1}{(1-q)}\lim_{\beta \rightarrow \infty}\left[\left(1 + \frac{(1-q)}{\beta}~S\right)^{\beta} - 1\right] 
 = \frac{1}{(1-q)}\left[\mathrm{e}^{(1-q)S} - 1\right]\label{LQG}
\end{eqnarray}
which resembles with the Loop Quantum Gravity entropy \cite{Majhi:2017zao,Liu:2021dvj}.

\end{itemize}

Furthermore, the generalized entropy function in Eq.~(\ref{gen-entropy}) shares the following properties: 
(1) $S_\mathrm{g}\left[ \alpha_{\pm},\beta,\gamma,\epsilon\right]$ tends to zero for $S \rightarrow 0$, i.e. the non-singular generalized entropy 
satisfies the generalized third law of thermodynamics. (2) Due to the fact that the hyperbolic terms present in the expression of $S_\mathrm{g}$ 
increase with $S$, the generalized entropy 
$S_\mathrm{g}\left[ \alpha_{\pm},\beta,\gamma,\epsilon \right]$ turns out to be a monotonically increasing function of $S$. 
(3) $S_\mathrm{g}\left[ \alpha_{\pm},\beta,\gamma,\epsilon \right]$ proves to converge to the Bekenstein-Hawking 
entropy at certain limit of the parameters. 
In particular, by taking $\epsilon \rightarrow 0$ and then $\alpha_+ \rightarrow \infty$, $\alpha_- = 0$, $\gamma = \left(\alpha_+/\beta\right)^{\beta}$ 
and $\beta = 1$, one can show that the generalized entropy function in Eq.~(\ref{gen-entropy}) becomes similar to the Bekenstein-Hawking entropy.

At this stage it deserves mentioning that we have proposed two different generalized entropy functions in Eq.(\ref{gen-entropy-earlier}) and 
in Eq.(\ref{gen-entropy}) respectively -- the former entropy function contains four independent parameters while the latter one 
has five parameters. Furthermore both the entropies are able to generalize the known entropies mentioned from Eq.(\ref{Tsallis-Barrow}) to 
Eq.(\ref{LQG}) for suitable choices of the respective parameters. However as mentioned earlier that the entropy with four parameters becomes singular 
at $H = 0$ (for instance, in a bounce scenario when the Hubble parameter vanishes at the instant of bounce), while the entropy function having five 
parameters proves to be singular-free during the whole cosmological evolution of the universe. Based on these findings, we give the following conjecture 
in regard to the non-singular generalized entropy function:\\

\textbf{\underline{A Conjecture}}: 
``The minimum number of parameters required in a generalized entropy function that can generalize 
all the known entropies mentioned from Eq.~(\ref{Tsallis-Barrow}) to Eq.~(\ref{LQG}), and at the same time, is also singular-free during the universe's evolution -- 
is equal to five''.

\section{Modified cosmology corresponding to the generalized entropy} \label{SecI}

In this section, we will address the cosmological field equations corresponds to the non-singular entropy $S_\mathrm{g}$ shown in Eq.(\ref{gen-entropy}). 
As we will show that the $S_\mathrm{g}$ introduces an effective energy density and an effective pressure into the 
Friedmann-Lema\^{i}tre-Robertson-Walker (FLRW) equations. 

The FLRW metric with flat spacial part will serve our purpose in the present context, i.e
\begin{align}
ds^2=-dt^2+a^2(t)\sum_{i=1,2,3} \left(dx^i\right)^2 \, ,
\label{metric}
\end{align}
where $t$ and $a(t)$ are cosmic time (or proper time for a comoving observer) and the scale factor of the universe respectively. 

The cosmological horizon has the radius given by, 
\begin{align}
\label{apphor}
r_\mathrm{H}=\frac{1}{H}\, ,
\end{align}
with $H = \dot{a}/a$ is known as the Hubble parameter of the universe. 
The amount of entropy within the cosmological horizon follows the Bekenstein-Hawking relation \cite{Padmanabhan:2009vy}. 
Moreover the flux of the energy $E$, or equivalently, the heat $Q$ within the cosmological horizon turns out to be
\begin{align}
\label{Tslls2}
dQ = - dE = -\frac{4\pi}{3} r_\mathrm{H}^3 \dot\rho dt = -\frac{4\pi}{3H^3} \dot\rho~dt 
= \frac{4\pi}{H^2} \left( \rho + p \right)~dt \, ,
\end{align}
where $\rho$ is the energy density of the normal matter under consideration, and we use the conservation law: $0 = \dot \rho + 3 H \left( \rho + p \right)$ in the last equality. Then the Hawking temperature \cite{Cai:2005ra}
\begin{align}
\label{Tslls6}
T = \frac{1}{2\pi r_\mathrm{H}} = \frac{H}{2\pi}\, ,
\end{align}
along with the first law of thermodynamics $TdS = dQ$ results to $\dot H = - 4\pi G \left( \rho + p \right)$ which is identical with the spatial 
part of the usual Friedmann equation. Integrating both sides of this equation (with respect to time) leads to the temporal part of the FRW equation, 
\begin{align}
\label{Tslls8}
H^2 = \frac{8\pi G}{3} \rho + \frac{\Lambda}{3} \, ,
\end{align}
where $\Lambda$ is the integration constant, and acts as a cosmological constant. 

We now apply the above formalism to the non-singular generalized entropy function $S_\mathrm{g}$, rather than the Bekenstein-Hawking entropy. In effect
of which, the first law of thermodynamics gives
\begin{eqnarray}
TdS_\mathrm{g} = dQ~~.
\label{N1}
\end{eqnarray}
Due to the fact that the generalized entropy is function of the Bekenstein-Hawking entropy, i.e $S_\mathrm{g} = S_\mathrm{g}(S)$ where $S$ is the Bekenstein-Hawking entropy, we may equivalently write Eq.(\ref{N1}) as $T\left(\frac{\partial S_\mathrm{g}}{\partial S}\right)dS = dQ$. By using Eq.(\ref{Tslls2}) and $S = \frac{\pi}{GH^2}$, the thermodynamic equation leads to the following evolution of the Hubble parameter as,
\begin{align}
\dot{H}\left(\frac{\partial S_\mathrm{g}}{\partial S}\right) = -4\pi G\left(\rho + p\right) \,,
\label{FRW1-sub}
\end{align}
which, due to the explicit form of $S_\mathrm{g}$ shown in Eq.(\ref{gen-entropy}), takes the following form,
\begin{eqnarray}
\frac{1}{\gamma}&\bigg[&\alpha_{+}~\mathrm{sech}^2\left(\frac{\epsilon\pi \alpha_+}{\beta GH^2}\right)
\left\{1 + \frac{1}{\epsilon}\tanh\left(\frac{\epsilon\pi \alpha_+}{\beta GH^2}\right)\right\}^{\beta-1}\nonumber\\
&+&\alpha_{-}~\mathrm{sech}^2\left(\frac{\epsilon\pi \alpha_-}{\beta GH^2}\right)
\left\{1 + \frac{1}{\epsilon}\tanh\left(\frac{\epsilon\pi \alpha_-}{\beta GH^2}\right)\right\}^{-\beta-1}\bigg]\dot{H} = -4\pi G\left(\rho + p\right)~~.
\label{FRW-1}
\end{eqnarray}
Owing to the conservation equation of matter fields, in particular $\dot{\rho} + 3H\left(\rho + p\right) = 0$, 
the above expression becomes,
\begin{eqnarray}
\frac{2}{\gamma}&\bigg[&\alpha_{+}~\mathrm{sech}^2\left(\frac{\epsilon\pi \alpha_+}{\beta GH^2}\right)
\left\{1 + \frac{1}{\epsilon}\tanh\left(\frac{\epsilon\pi \alpha_+}{\beta GH^2}\right)\right\}^{\beta-1}\nonumber\\
&+&\alpha_{-}~\mathrm{sech}^2\left(\frac{\epsilon\pi \alpha_-}{\beta GH^2}\right)
\left\{1 + \frac{1}{\epsilon}\tanh\left(\frac{\epsilon\pi \alpha_-}{\beta GH^2}\right)\right\}^{-\beta-1}\bigg]H~dH = \left(\frac{8\pi G}{3}\right)d\rho \,,
 \nonumber
\end{eqnarray}
which, by integration on both sides, 
\begin{align}
f\left(H;~\alpha_{\pm},\beta,\gamma,\epsilon\right) = \frac{8\pi G\rho}{3} + \frac{\Lambda}{3} \,.
\label{FRW-2}
\end{align}
Here the integration constant is symbolized by $\Lambda$ and the function $f$ has the following form:
\begin{eqnarray}
 f\left(H;~\alpha_{\pm},\beta,\gamma,\epsilon\right) = 
 \frac{2}{\gamma}\int&\bigg[&\alpha_{+}~\mathrm{sech}^2\left(\frac{\epsilon\pi \alpha_+}{\beta GH^2}\right)
\left\{1 + \frac{1}{\epsilon}\tanh\left(\frac{\epsilon\pi \alpha_+}{\beta GH^2}\right)\right\}^{\beta-1}\nonumber\\
&+&\alpha_{-}~\mathrm{sech}^2\left(\frac{\epsilon\pi \alpha_-}{\beta GH^2}\right)
\left\{1 + \frac{1}{\epsilon}\tanh\left(\frac{\epsilon\pi \alpha_-}{\beta GH^2}\right)\right\}^{-\beta-1}\bigg]H~dH~~.
\label{f}
\end{eqnarray}
In regard to the functional form of $f\left(H;~\alpha_{\pm},\beta,\gamma,\epsilon\right)$, we would like to mention that the integration 
in Eq.(\ref{f}) may not be performed in a closed form, unless certain conditions are imposed. 
For example, we consider $GH^2 \ll 1$ which is, in fact, valid during the 
universe's evolution (i.e the Hubble parameter is less than the Planck scale). With $GH^2 \ll 1$, Eq.(\ref{f}) becomes,
\begin{eqnarray}
 f\left(H;~\alpha_{\pm},\beta,\gamma,\epsilon\right) = 
 \frac{8}{\gamma}\int\left\{\alpha_{+}\left(1+\frac{1}{\epsilon}\right)^{\beta-1}\mathrm{exp}\left(-\frac{2\epsilon\pi \alpha_+}{\beta GH^2}\right)
 + \alpha_{-}\left(1+\frac{1}{\epsilon}\right)^{-\beta-1}\mathrm{exp}\left(-\frac{2\epsilon\pi \alpha_-}{\beta GH^2}\right)\right\}H~dH~~,
 \label{f-aprroximated-1}
\end{eqnarray}
which can be integrated, and consequently, we get the following form of $f$:
\begin{eqnarray}
 f\left(H;~\alpha_{\pm},\beta,\gamma,\epsilon\right) = 
 \frac{4}{\gamma}H^2&\bigg\{&\alpha_{+}\left(1+\frac{1}{\epsilon}\right)^{\beta-1}\left[\mathrm{exp}\left(-\frac{2\epsilon\pi \alpha_+}{\beta GH^2}\right) 
 + \left(\frac{2\epsilon\pi \alpha_+}{\beta GH^2}\right)\mathrm{Ei}\left(-\frac{2\epsilon\pi \alpha_+}{\beta GH^2}\right)\right]\nonumber\\
 &+&\alpha_{-}\left(1+\frac{1}{\epsilon}\right)^{-\beta-1}\left[\mathrm{exp}\left(-\frac{2\epsilon\pi \alpha_-}{\beta GH^2}\right) 
 + \left(\frac{2\epsilon\pi \alpha_-}{\beta GH^2}\right)\mathrm{Ei}\left(-\frac{2\epsilon\pi \alpha_-}{\beta GH^2}\right)\right]\bigg\}~~.
 \label{f-aprroximated-2}
\end{eqnarray}
Therefore as a whole, the general form of $f\left(H;~\alpha_{\pm},\beta,\gamma,\epsilon\right)$ is given in Eq.(\ref{f}). However with the 
consideration of $GH^2 \ll 1$, the integration of Eq.(\ref{f}) is performed and hence 
a closed form of $f\left(H;~\alpha_{\pm},\beta,\gamma,\epsilon\right)$ is obtained in Eq.(\ref{f-aprroximated-2}).

Eq.~(\ref{FRW-1}) and Eq.~(\ref{FRW-2}) are the cosmological field equations 
corresponding to the generalized entropy $S_\mathrm{g}$. The presence of the entropy $S_\mathrm{g}$ effectively produces an 
energy density and pressure in the modified Friedmann equations. To be more explicit, we now define the following energy density and pressure,
\begin{align}
\rho_\mathrm{g} = \frac{3}{8\pi G}\left\{ H^2 - f\left(H;~\alpha_{\pm},\beta,\gamma,\epsilon\right) \right\} \,,
\label{efective energy density}
\end{align}
and
\begin{eqnarray}
p_\mathrm{g} = \frac{\dot{H}}{4\pi G}\bigg\{\frac{1}{\gamma}&\bigg[&\alpha_{+}~\mathrm{sech}^2\left(\frac{\epsilon\pi \alpha_+}{\beta GH^2}\right)
\left\{1 + \frac{1}{\epsilon}\tanh\left(\frac{\epsilon\pi \alpha_+}{\beta GH^2}\right)\right\}^{\beta-1}\nonumber\\
&+&\alpha_{-}~\mathrm{sech}^2\left(\frac{\epsilon\pi \alpha_-}{\beta GH^2}\right)
\left\{1 + \frac{1}{\epsilon}\tanh\left(\frac{\epsilon\pi \alpha_-}{\beta GH^2}\right)\right\}^{-\beta-1}\bigg] - 1\bigg\} - \rho_\mathrm{g} \,,
\label{effecyive pressure}
\end{eqnarray}
respectively. As a consequence, Eq.~(\ref{FRW-1}) and Eq.~(\ref{FRW-2}) can be equivalently expressed as, 
\begin{align}
\dot{H}=&\,-4\pi G\left[\left(\rho + \rho_\mathrm{g}\right) + \left(p + p_\mathrm{g}\right)\right] \,,\nonumber\\
H^2=&\, \frac{8\pi G}{3}\left(\rho + \rho_\mathrm{g}\right) + \frac{\Lambda}{3} \,.
\label{final FRW}
\end{align}
This demonstrates that Eq.(\ref{FRW-1}) and Eq.(\ref{FRW-2}) are similar to the usual Friedmann equations with the total energy density and pressure 
are given by $\rho_\mathrm{T} = \rho + \rho_\mathrm{g}$ and $p_\mathrm{T} = p + p_\mathrm{g}$ respectively. Therefore $\rho_\mathrm{g}$ and 
$p_\mathrm{g}$ denote the energy density and pressure produced by the non-singular generalized entropy $S_\mathrm{g}$ itself. The motivation of the 
present paper is to investigate the possible implications of $\rho_\mathrm{g}$ and $p_\mathrm{g}$ on bounce cosmology. In particular, we will show that 
the modified Friedmann Eq.(\ref{final FRW}) naturally allows a non-singular universe. 

However, before moving to such bounce scenario, we will show that the entropic cosmology corresponds to 
the $S_\mathrm{g}$ can be regarded to be equivalent to the generalized holographic cosmology with suitable cut-off.

\section{Equivalence between the entropic cosmology and the generalized holographic cosmology}\label{sec-equivalence}

The holographic energy density, in the realm of holographic principle, comes as,
\begin{align}
\label{basic}
\rho_\mathrm{hol}=\frac{3c^2}{\kappa^2 L^2_\mathrm{IR}}\, ,
\end{align}
where $L_\mathrm{IR}$ is known as the infrared cut-off, $c$ is a free parameter and $\kappa^2 = 8\pi G$ with $G$ being the gravitational constant. 
Here the particle horizon (symbolized by $L_\mathrm{p}$) or the future event horizon
(symbolized by $L_\mathrm{f}$) are defined as,
\begin{align}
\label{H3}
L_\mathrm{p}\equiv a \int_0^t\frac{dt}{a} \,,\quad
L_\mathrm{f}\equiv a \int_t^\infty \frac{dt}{a}\, .
\end{align}
A differentiation (with respect to $t$) of both sides of the above expressions yields 
the Hubble parameter in terms of particle horizon and its derivative or in terms of future horizon and its derivative as,
\begin{align}
\label{HLL}
H \left( L_\mathrm{p} , \dot{L}_\mathrm{p} \right) = \frac{\dot{L}_\mathrm{p}}{L_\mathrm{p}} - \frac{1}{L_\mathrm{p}}\, , 
\quad H(L_\mathrm{f} , \dot{L}_\mathrm{f}) = \frac{\dot{L}_\mathrm{f}}{L_\mathrm{f}} + \frac{1}{L_\mathrm{f}} \, .
\end{align}
In regard to the holographic cut-off, a general form was proposed in \cite{Nojiri:2005pu} as,
\begin{align}
\label{GeneralLIR}
L_\mathrm{IR} = L_\mathrm{IR} \left( L_\mathrm{p}, \dot L_\mathrm{p}, 
\ddot L_\mathrm{p}, \cdots, L_\mathrm{f}, \dot L_\mathrm{f}, \cdots, a\right)\, .
\end{align}
It may be observed that $L_\mathrm{IR}$ depends on $L_\mathrm{p}$, $L_\mathrm{f}$ and their derivatives, and the scale factor. 
The other dependency of $L_\mathrm{IR}$, particularly on the Ricci scalar and its derivatives, are embedded 
by either $L_\mathrm{p}$, $\dot{L}_\mathrm{p}$ or $L_\mathrm{f}$, $\dot{L}_\mathrm{f}$ via Eq.~(\ref{HLL}). 
Such a generalized cutoff may correspond to a general covariant gravity model,
\begin{align}
\label{GeneralAc}
S = \int d^4 \sqrt{-g} F \left( R,R_{\mu\nu} R^{\mu\nu},
R_{\mu\nu\rho\sigma}R^{\mu\nu\rho\sigma}, \Box R, \Box^{-1} R,
\nabla_\mu R \nabla^\mu R, \cdots \right) \, .
\end{align}
Here it deserves mentioning that all the HDE models proposed so far (for example the Tsallis HDE or the Barrow HDE etc.) 
are shown to be different candidates of generalized holographic cosmology, see \cite{Nojiri:2021iko,Nojiri:2021jxf}. In this section, 
we will examine whether the entropic cosmology corresponds to the present entropy function 
$S_\mathrm{g}$ is equivalent to the generalized holographic cosmology with specific cut-offs.

Using Eq.~(\ref{efective energy density}) and Eq.~(\ref{basic}), we may argue that the entropic energy density can be thought to be equivalent 
with the generalized holographic energy density, where the equivalent holographic cutoff $L_\mathrm{g}$ depends on either the particle horizon and its 
derivative or the future horizon and its derivative. In the former case, $L_\mathrm{g}$ is given by, 
\begin{eqnarray}
\label{e-1}
\frac{3c^2}{\kappa^2 L^2_\mathrm{g}} 
= \frac{3}{8\pi G} 
\left\{\left( \frac{\dot{L}_\mathrm{p}}{L_\mathrm{p}} - \frac{1}{L_\mathrm{p}} \right)^2 
- f_1\left(L_\mathrm{p},\dot{L}_\mathrm{p}\right)\right\}~~,
\end{eqnarray}
in terms of $L_\mathrm{p}$, $\dot{L}_\mathrm{p}$. Here $f_1\left(L_\mathrm{p},\dot{L}_\mathrm{p} \right)$ has the following 
form:
\begin{eqnarray}
f_1\left(L_\mathrm{p},\dot{L}_\mathrm{p} \right) 
= \frac{2}{\gamma}\int&\bigg[&\alpha_{+}~\mathrm{sech}^2\left(\frac{\epsilon\pi \alpha_+}{\beta GH^2}\right)
\left\{1 + \frac{1}{\epsilon}\tanh\left(\frac{\epsilon\pi \alpha_+}{\beta GH^2}\right)\right\}^{\beta-1}\nonumber\\
&+&\alpha_{-}~\mathrm{sech}^2\left(\frac{\epsilon\pi \alpha_-}{\beta GH^2}\right)
\left\{1 + \frac{1}{\epsilon}\tanh\left(\frac{\epsilon\pi \alpha_-}{\beta GH^2}\right)\right\}^{-\beta-1}\bigg]H~dH\bigg|_
{H = \frac{\dot{L}_\mathrm{p}}{L_\mathrm{p}} - \frac{1}{L_\mathrm{p}}}~~.
\label{e-1-1}
\end{eqnarray}
Similarly, $L_\mathrm{g}$ in terms of the future horizon and its derivative becomes,
\begin{eqnarray}
\label{e-2}
\frac{3c^2}{\kappa^2 L^2_\mathrm{g}} 
= \frac{3}{8\pi G} 
\left\{\left( \frac{\dot{L}_\mathrm{f}}{L_\mathrm{f}} + \frac{1}{L_\mathrm{f}} \right)^2 
- f_2\left(L_\mathrm{f},\dot{L}_\mathrm{f}\right)\right\}~~,
\end{eqnarray}
with $f_2\left(L_\mathrm{f},\dot{L}_\mathrm{f} \right)$ is given by, 
\begin{eqnarray}
f_2\left(L_\mathrm{p},\dot{L}_\mathrm{p} \right) 
= \frac{2}{\gamma}\int&\bigg[&\alpha_{+}~\mathrm{sech}^2\left(\frac{\epsilon\pi \alpha_+}{\beta GH^2}\right)
\left\{1 + \frac{1}{\epsilon}\tanh\left(\frac{\epsilon\pi \alpha_+}{\beta GH^2}\right)\right\}^{\beta-1}\nonumber\\
&+&\alpha_{-}~\mathrm{sech}^2\left(\frac{\epsilon\pi \alpha_-}{\beta GH^2}\right)
\left\{1 + \frac{1}{\epsilon}\tanh\left(\frac{\epsilon\pi \alpha_-}{\beta GH^2}\right)\right\}^{-\beta-1}\bigg]H~dH\bigg|_
{H = \frac{\dot{L}_\mathrm{f}}{L_\mathrm{f}} + \frac{1}{L_\mathrm{f}}}~~.
\label{e-2-1}
\end{eqnarray}
With the condition $GH^2 \ll 1$ along with Eq.(\ref{f-aprroximated-2}), the integral in Eq.(\ref{e-1-1}) (or in Eq.(\ref{e-2-1})) is performed, 
and thus the $L_\mathrm{g}$ can be achieved in a closed form as,
\begin{eqnarray}
 \frac{3c^2}{\kappa^2 L^2_\mathrm{g}} 
&=&\frac{3}{8\pi G}\left( \frac{\dot{L}_\mathrm{p}}{L_\mathrm{p}} - \frac{1}{L_\mathrm{p}} \right)^2
\Bigg[1 - \frac{4}{\gamma}\bigg\{\alpha_{+}\left(1+\frac{1}{\epsilon}\right)^{\beta-1}\left(\mathrm{exp}\left(-\frac{2\epsilon\pi \alpha_+}{\beta GH^2}\right) 
 + \left(\frac{2\epsilon\pi \alpha_+}{\beta GH^2}\right)\mathrm{Ei}\left(-\frac{2\epsilon\pi \alpha_+}{\beta GH^2}\right)\right)\nonumber\\
 &+&\alpha_{-}\left(1+\frac{1}{\epsilon}\right)^{-\beta-1}\left(\mathrm{exp}\left(-\frac{2\epsilon\pi \alpha_-}{\beta GH^2}\right) 
 + \left(\frac{2\epsilon\pi \alpha_-}{\beta GH^2}\right)\mathrm{Ei}\left(-\frac{2\epsilon\pi \alpha_-}{\beta GH^2}\right)\right)\bigg\}\Bigg]\bigg|_
{H = \frac{\dot{L}_\mathrm{p}}{L_\mathrm{p}} - \frac{1}{L_\mathrm{p}}}
\label{e-1 approximated}
\end{eqnarray}
in terms of $L_\mathrm{p}$ and $\dot{L}_\mathrm{p}$, or,
\begin{eqnarray}
 \frac{3c^2}{\kappa^2 L^2_\mathrm{g}} 
&=&\frac{3}{8\pi G}\left( \frac{\dot{L}_\mathrm{f}}{L_\mathrm{f}} + \frac{1}{L_\mathrm{f}} \right)^2
\Bigg[1 - \frac{4}{\gamma}\bigg\{\alpha_{+}\left(1+\frac{1}{\epsilon}\right)^{\beta-1}\left(\mathrm{exp}\left(-\frac{2\epsilon\pi \alpha_+}{\beta GH^2}\right) 
 + \left(\frac{2\epsilon\pi \alpha_+}{\beta GH^2}\right)\mathrm{Ei}\left(-\frac{2\epsilon\pi \alpha_+}{\beta GH^2}\right)\right)\nonumber\\
 &+&\alpha_{-}\left(1+\frac{1}{\epsilon}\right)^{-\beta-1}\left(\mathrm{exp}\left(-\frac{2\epsilon\pi \alpha_-}{\beta GH^2}\right) 
 + \left(\frac{2\epsilon\pi \alpha_-}{\beta GH^2}\right)\mathrm{Ei}\left(-\frac{2\epsilon\pi \alpha_-}{\beta GH^2}\right)\right)\bigg\}\Bigg]\bigg|_
{H = \frac{\dot{L}_\mathrm{f}}{L_\mathrm{f}} - \frac{1}{L_\mathrm{f}}}
\label{e-2 approximated}
\end{eqnarray}
in terms of $L_\mathrm{f}$ and $\dot{L}_\mathrm{f}$. 

Wer now intend to determine the equation of state (EoS) parameter of $\rho_\mathrm{hol} = 3c^2/\left(\kappa^2L_\mathrm{g}^2\right)$, i.e for 
the holographic energy density with the cut-off given by $L_\mathrm{g}$. 
In effect of the conservation relation of $\rho_\mathrm{hol}$, one may write the corresponding EoS parameter ($\Omega_\mathrm{hol}^{(g)}$ where 
the superscript '$\mathrm{g}$' denotes that the EoS parameter corresponds to the cut-off $L_\mathrm{g}$) as, 
\begin{align}
\Omega_\mathrm{hol}^{(g)} = -1 - \left(\frac{2}{3HL_\mathrm{g}}\right)\frac{dL_\mathrm{g}}{dt} \,,
\label{eos-holg}
\end{align}
where $L_\mathrm{g}$ is shown in Eq.~(\ref{e-1}) (or in Eq.~(\ref{e-2})). 
Owing to Eq.~(\ref{HLL}), the above form of $\Omega_\mathrm{hol}^{(g)}$ turns out to be 
equivalent with $\omega_\mathrm{g} = p_\mathrm{g}/\rho_\mathrm{g}$, i.e. the EoS parameter corresponds to the holographic energy density 
is equivalent with that of corresponds to the entropic energy density. In particular,
\begin{align}
\Omega_\mathrm{hol}^{(g)} \equiv \omega_{g} \,.
\label{e-3}
\end{align}
Therefore we may argue that the entropic cosmology corresponds to the non-singular entropy $S_\mathrm{g}$ can be thought as a candidate 
of the generalized holographic family where the corresponding holographic cut-off is represented in terms of either $L_\mathrm{p}$ and $\dot{L}_\mathrm{p}$ (see
Eq.~(\ref{e-1})) or in  terms of $L_\mathrm{f}$ and $\dot{L}_\mathrm{f}$ (see Eq.~(\ref{e-2})).

\section{Generalized entropy on bounce cosmology}\label{sec-bounce}

For the first time, we provide a non-singular generalized entropy ($S_\mathrm{g}$), in particular, all the known entropies proposed so far (like 
Tsallis, Barrow, R\'{e}nyi, Sharma-Mittal, 
Kaniadakis and Loop Quantum Gravity entropies) become singular (or diverge) 
when the Hubble parameter vanishes during the universe's evolution (for instance, in bounce cosmology at the instant of bounce), 
unlike to the $S_\mathrm{g}[\alpha_{\pm},\beta,\gamma,\epsilon]$ which proves to be singular-free at $H = 0$. 
Such $non-singular$ behaviour of the proposed entropy function is useful in describing bouncing scenario, in which case, the universe undergoes through 
$H = 0$ at the instant of bounce. 

In this section, we will address the implications of the generalized entropy $S_\mathrm{g}$ on non-singular bounce cosmology, in particular, 
we will investigate whether the entropic energy density can trigger a viable bounce during the early stage of the universe that is consistent 
with the observational constraints. For this purpose, we take 
the matter field and the cosmological constant to be absent, i.e., $\rho = p = \Lambda = 0$. In effect, Eq.~(\ref{FRW-1}) and 
Eq.~(\ref{FRW-2}) becomes,
\begin{eqnarray}
\frac{1}{\gamma}&\bigg[&\alpha_{+}~\mathrm{sech}^2\left(\frac{\epsilon\pi \alpha_+}{\beta GH^2}\right)
\left\{1 + \frac{1}{\epsilon}\tanh\left(\frac{\epsilon\pi \alpha_+}{\beta GH^2}\right)\right\}^{\beta-1}\nonumber\\
&+&\alpha_{-}~\mathrm{sech}^2\left(\frac{\epsilon\pi \alpha_-}{\beta GH^2}\right)
\left\{1 + \frac{1}{\epsilon}\tanh\left(\frac{\epsilon\pi \alpha_-}{\beta GH^2}\right)\right\}^{-\beta-1}\bigg]\dot{H} = 0~~.
\label{FRW-1-bounce}
\end{eqnarray}
The parameters $\left(\alpha_{\pm},\beta,\gamma,\epsilon\right)$ are positive, and thus the solution of the above equation is given by: 
$\dot{H} = 0$ or equivalently $H=\mathrm{constant}$. Therefore the cosmology corresponds to the generalized entropy $S_\mathrm{g}[\alpha_{\pm}
,\beta,\gamma,\epsilon]$ results to a constant Hubble parameter of the universe. Here we would like to mention that the emergence of a constant 
Hubble parameter is a generic property for all the known entropy functions, like the Tsallis, the R\'{e}nyi, the Kaniadakis entropy etc. Clearly
$H=\mathrm{constant}$ does not lead to the correct evolution of the universe. Thus in order to have an acceptable cosmological evolution in the present 
context, we consider the parameters of $S_\mathrm{g}[\alpha_{\pm},\beta,\gamma,\epsilon]$ vary with time 
(see \cite{Nojiri:2022dkr,Nojiri:2019skr} where the entropic cosmology with $varying$ exponents were studied). 
The running behavior of such parameters 
may be motivated by quantum gravity, particularly in the case of gravity, if the space-time fluctuates at high energy scales, 
the degrees of freedom may increase. On the other hand, if gravity becomes a topological theory, the degrees of freedom may decrease. 
In particular, we consider the parameter $\gamma$ to vary with time, and all the other parameters remain fixed, i.e.
\begin{eqnarray}
 \gamma = \gamma(N)~~,
 \label{gamma}
\end{eqnarray}
with $N$ being the e-fold number of the universe. In such scenario where $\gamma(N)$ varies with time, the Friedmann equation corresponds to 
$S_\mathrm{g}[\alpha_{\pm},\beta,\gamma,\epsilon]$ gets modified compared to Eq.(\ref{FRW-1-bounce}), and is given by:
\begin{eqnarray}
 \left(\frac{2\pi}{G}\right)\left(\frac{\partial S_\mathrm{g}}{\partial S}\right)\frac{H'(N)}{H^3}  = 
 \left(\frac{\partial S_\mathrm{g}}{\partial \gamma}\right)\gamma'(N)~~,
 \label{bounce-1}
\end{eqnarray}
where the overprime denotes the derivative with respect to $N$. With the explicit form of $S_\mathrm{g}$, Eq.(\ref{bounce-1}) takes the following form,
\begin{eqnarray}
 &-&\left(\frac{2\pi}{G}\right)\left(\frac{H'(N)}{H^3}\right)\times\nonumber\\
 &\Bigg[&\frac{\alpha_{+}~\mathrm{sech}^2\left(\frac{\epsilon \alpha_+}{\beta}S\right)
\left\{1 + \frac{1}{\epsilon}\tanh\left(\frac{\epsilon\alpha_+}{\beta}S\right)\right\}^{\beta-1} 
+ \alpha_{-}~\mathrm{sech}^2\left(\frac{\epsilon\alpha_-}{\beta}S\right)
\left\{1 + \frac{1}{\epsilon}\tanh\left(\frac{\epsilon\alpha_-}{\beta}S\right)\right\}^{-\beta-1}}
{\left\{1 + \frac{1}{\epsilon}\tanh\left(\frac{\epsilon\alpha_+}{\beta}S\right)\right\}^{\beta} - 
\left\{1 + \frac{1}{\epsilon}\tanh\left(\frac{\epsilon\alpha_-}{\beta}S\right)\right\}^{-\beta}}\Bigg] = \frac{\gamma'(N)}{\gamma(N)}~.
\label{bounce-2}
\end{eqnarray}
The above equation clearly depicts that due to $\gamma'(N) \neq 0$, the Hubble parameter is not a constant in this context, and thus 
it may lead to a viable non-singular bounce. Due to $S = \pi/(GH^2)$, Eq.(\ref{bounce-2}) turns out to be,
\begin{eqnarray}
\left[\frac{\alpha_{+}~\mathrm{sech}^2\left(\frac{\epsilon \alpha_+}{\beta}S\right)
\left\{1 + \frac{1}{\epsilon}\tanh\left(\frac{\epsilon\alpha_+}{\beta}S\right)\right\}^{\beta-1} 
+ \alpha_{-}~\mathrm{sech}^2\left(\frac{\epsilon\alpha_-}{\beta}S\right)
\left\{1 + \frac{1}{\epsilon}\tanh\left(\frac{\epsilon\alpha_-}{\beta}S\right)\right\}^{-\beta-1}}
{\left\{1 + \frac{1}{\epsilon}\tanh\left(\frac{\epsilon\alpha_+}{\beta}S\right)\right\}^{\beta} - 
\left\{1 + \frac{1}{\epsilon}\tanh\left(\frac{\epsilon\alpha_-}{\beta}S\right)\right\}^{-\beta}}\right]dS = \frac{\gamma'(N)}{\gamma(N)}dN
\label{bounce-3}
\end{eqnarray}
which can be integrated to get,
\begin{eqnarray}
 \left\{1 + \frac{1}{\epsilon}\tanh\left(\frac{\epsilon \pi\alpha_+}{\beta GH^2}\right)\right\}^{\beta} - 
\left\{1 + \frac{1}{\epsilon}\tanh\left(\frac{\epsilon \pi\alpha_-}{\beta GH^2}\right)\right\}^{-\beta} = \gamma(N)~~.
\label{bounce-4}
\end{eqnarray}
The above equation provides the Hubble parameter in terms of e-fold number, i.e. $H = H(N)$, for a suitable form of $\gamma(N)$. 
In order to extract an explicit form of the Hubble parameter from Eq.(\ref{bounce-4}), we take $\alpha_+ = \alpha_- = \alpha$ (say) 
without losing any generality. In effect, Eq.(\ref{bounce-4}) yields $H=H(N)$ as,
\begin{eqnarray}
 \tanh{\left(\frac{\epsilon \pi\alpha}{\beta GH^2}\right)} = \left\{\frac{\gamma(N) + \sqrt{\gamma^2(N) + 4}}{2}\right\}^{1/\beta} - 1~~.
 \label{bounce-6}
\end{eqnarray}
Due to the appearance of quadratic power of $H$, Eq.(\ref{bounce-6}) allows a positive branch as well as a negative branch of the Hubble parameter. 
This leads to a natural possibility of symmetric bounce in the present context of singular free generalized entropic cosmology. Moreover Eq.(\ref{bounce-6}) 
also demonstrates that the explicit evolution of $H(N)$ does depend on the form of $\gamma(N)$. In the following, we will consider two cases where 
we will determine the form of $\gamma(N)$ such that it gives two different symmetric bounce scenarios respectively. However before examining 
the possibility of bounce scenarios, here we provide the effective energy density ($\rho_\mathrm{eff}$) and the effective pressure 
($p_\mathrm{eff}$) sourced from the $S_\mathrm{g}$ where 
the parameter $\gamma$ varies with the e-folding number. In particular, Eq.(\ref{bounce-2}) and Eq.(\ref{bounce-6}) immediately lead to the following 
forms of $\rho_\mathrm{eff}$ and $p_\mathrm{eff}$ as,
\begin{eqnarray}
 \rho_\mathrm{eff}&=&\left(\frac{3\epsilon\alpha}{4\beta G^2}\right)
 \left[\ln{\left\{\frac{1}{2\left(\frac{2}{\gamma(N) + \sqrt{\gamma^2(N) + 4}}\right)^{1/\beta} - 1}\right\}}\right]^{-1}~~,\nonumber\\
 p_\mathrm{eff} + \rho_\mathrm{eff}&=&\left(\frac{\gamma'(N)}{8\pi^2\gamma(N)}\right)
 \left[\frac{H^4\left[\left\{1 + \frac{1}{\epsilon}\tanh\left(\frac{\epsilon\pi\alpha}{\beta GH^2}\right)\right\}^{\beta} - 
\left\{1 + \frac{1}{\epsilon}\tanh\left(\frac{\epsilon\pi\alpha}{\beta GH^2}\right)\right\}^{-\beta}\right]}
{\alpha~\mathrm{sech}^2\left(\frac{\epsilon \pi\alpha}{\beta GH^2}\right)\left[
\left\{1 + \frac{1}{\epsilon}\tanh\left(\frac{\epsilon \pi\alpha}{\beta GH^2}\right)\right\}^{\beta-1} 
+ \left\{1 + \frac{1}{\epsilon}\tanh\left(\frac{\epsilon \pi\alpha}{\beta GH^2}\right)\right\}^{-\beta-1}\right]}\right]\nonumber\\
\label{energy density and pressure}
\end{eqnarray}
respectively. These expressions will be useful later.

\subsection{Possibility for an exponential bounce}
Here the scale factor is taken as,
\begin{eqnarray}
 a(t) = \mathrm{exp}\left(a_0t^2\right)
 \label{exp bounce-1}
\end{eqnarray}
which results to a symmetric bounce at $t = 0$. Here $a_0$ is a constant having mass dimension [+2] -- this constant is related 
with the entropic parameters of $S_\mathrm{g}$ and thus, without losing any generality, 
we take $a_0 = \frac{\epsilon \pi\alpha}{4G\beta}$. Hence the scale factor takes the following form:
\begin{eqnarray}
 a(t) = \mathrm{exp}\left(\frac{\epsilon \pi\alpha}{4G\beta}~t^2\right)~~.
 \label{exp bounce-2}
\end{eqnarray}
Consequently the Hubble parameter in terms of e-fold number comes as,
\begin{eqnarray}
 H(N) = \pm \sqrt{\frac{\epsilon \pi\alpha}{G\beta}}~N^{1/2}~~,
 \label{exp bounce-3}
\end{eqnarray}
where we use the relation between the cosmic time and the e-fold number $t(N) = \pm \sqrt{\frac{4G\beta}{\epsilon \pi\alpha}}N^{\frac{1}{2}}$ obtained 
from $N = \ln{a}$. Eq.(\ref{exp bounce-3}) clearly indicates that the Hubble parameter is negative in the negative branch of $t(N)$, while $H(N) > 0$ 
at $t(N) > 0$. Therefore the negative and positive branch of $t(N)$ represents the contracting and expanding stage of the universe respectively. 
Clearly the bounce occurs at $t = 0$ or equivalently at $N = 0$. Thus in terms of the e-fold number, the universe starts at $N \rightarrow +\infty$ 
from the distant past, then the bounce happens at $N = 0$ and consequently the universe goes to the distant future again at $N \rightarrow +\infty$. 
By using Eq.(\ref{bounce-6}), we reconstruct the form of $\gamma(N)$ which allows the above $H = H(N)$, and is given by,
\begin{eqnarray}
 \gamma(N) = \left\{1 + \frac{1}{\epsilon}\tanh\left(\frac{1}{N}\right)\right\}^{\beta} - 
\left\{1 + \frac{1}{\epsilon}\tanh\left(\frac{1}{N}\right)\right\}^{-\beta}~~.
\label{exp bounce-4}
\end{eqnarray}
Therefore the exponential bounce scenario described by the scale factor in Eq.(\ref{exp bounce-2}) or equivalently by the Hubble parameter in 
Eq.(\ref{exp bounce-3}) can be achieved from the singular free generalized entropy $S_\mathrm{g}\left[\alpha,\beta,\epsilon,\gamma(N)\right]$ 
with $\gamma(N)$ is given by Eq.(\ref{exp bounce-4}). 

At this stage it deserves mentioning that in the case of exponential bounce, the comoving Hubble radius ( defined by $r_\mathrm{h} = 1/\left|aH\right|$, 
where $r_\mathrm{h}$ symbolizes the comoving Hubble radius ) decreases with time and asymptotically goes to zero at both sides of the bounce. 
This indicates that the perturbation modes generate near the bounce when the comoving Hubble radius is infinite in size to contain 
all the perturbation modes within the sub-Hubble regime. In effect, the issue of horizon problem appears as the perturbation modes 
are in the super-Hubble regime at the distant past. Due to such problem in the exponential bounce, we now consider an alternative bounce in the present context 
of entropic cosmology, which is free from the horizon problem. 

\subsection{Possibility for a quasi matter bounce}
In this case, the scale factor is,
\begin{eqnarray}
 a(t) = \left[1 + a_0\left(\frac{t}{t_0}\right)^2\right]^n
 \label{matter bounce-1}
\end{eqnarray}
which is symmetric about $t = 0$ when the bounce happens. The $n$, $a_0$ and $t_0$ considered in the scale factor are related to 
the entropic parameters, and we take it as follows:
\begin{eqnarray}
 n = \sqrt{\alpha}~~~~~~,~~~~~~~~a_0 = \frac{\pi}{4\beta}~~~~~~~~\mathrm{and}~~~~~~~~~t_0 = \sqrt{G/\epsilon}~~,
 \label{matter bounce-2}
\end{eqnarray}
with $G$ being the gravitational constant. The relation between ($n$, $a_0$, $t_0$) with the entropic parameters can be considered in a different 
way compared to the Eq.(\ref{matter bounce-2}), however for a simplified expression of $\gamma(N)$ we consider the relations as of 
Eq.(\ref{matter bounce-2}). Consequently the scale factor has the following form:
\begin{eqnarray}
 a(t) = \left[1 + \left(\frac{\pi\epsilon}{4\beta G}\right)t^2\right]^{\sqrt{\alpha}}~~.
 \label{matter bounce-3}
\end{eqnarray}
For $\alpha = \frac{1}{9}$, the scale factor represents a matter bounce scenario, while a quasi matter bounce is depicted by $\alpha \approx \frac{1}{9}$. 
At this stage, it deserves mentioning that due to the above scale factor, the comoving Hubble radius asymptotically goes as 
$r_h \sim t^{1-2\sqrt{\alpha}}$. Therefore for $\alpha < \frac{1}{4}$, the comoving Hubble radius asymptotically diverges to infinity, and hence, 
the primordial perturbation modes generate far before the bounce during the contracting phase. This results to the resolution of the horizon 
problem as the perturbation modes lie within the sub-Hubble domain at the distant past. However for $\alpha > \frac{1}{4}$, similar to the 
exponential bounce, $r_h$ asymptotically vanishes and consequently the bounce scenario may suffer from the Horizon issue. Based on these 
arguments, we will take $\alpha < \frac{1}{4}$ for the scale factor of Eq.(\ref{matter bounce-3}), which covers 
the range required for the quasi matter bounce. 

Eq.(\ref{matter bounce-3}) immediately gives the cosmic time in terms of e-fold number as follows:
\begin{eqnarray}
 t(N) = \pm\sqrt{\frac{4\beta G}{\pi\epsilon}}\left[\mathrm{e}^{N/\sqrt{\alpha}} - 1\right]^{\frac{1}{2}}~~.
 \label{matter bounce-4}
\end{eqnarray}
We use the above expression to get the Hubble parameter in terms of e-fold number, and is given by,
\begin{eqnarray}
 H(N) = \pm\left(\sqrt{\frac{\epsilon \pi\alpha}{\beta G}}\right)\mathrm{e}^{-N/\sqrt{\alpha}}\left[\mathrm{e}^{N/\sqrt{\alpha}} - 1\right]^{\frac{1}{2}}~~.
 \label{matter bounce-5}
\end{eqnarray}
Plugging back the above expression of $H(N)$ into Eq.(\ref{bounce-6}) yields the respective form of $\gamma(N)$ as,
\begin{eqnarray}
\gamma(N) = \left\{1 + \frac{1}{\epsilon}\tanh\left[\mathrm{e}^{-N/\sqrt{\alpha}}\left(\mathrm{e}^{N/\sqrt{\alpha}} - 1\right)^{\frac{1}{2}}\right]\right\}^{\beta} - 
\left\{1 + \frac{1}{\epsilon}\tanh\left[\mathrm{e}^{-N/\sqrt{\alpha}}\left(\mathrm{e}^{N/\sqrt{\alpha}} - 1\right)^{\frac{1}{2}}\right]\right\}^{-\beta}~~,
\label{matter bounce-6}
\end{eqnarray}
which triggers the quasi matter bounce described by the scale factor of Eq.(\ref{matter bounce-3}) in the present context of 
singular free generalized entropic cosmology.

Having described the background evolution, we now focus to the perturbation analysis for the curvature perturbation as well as 
for the tensor perturbation respectively. As mentioned earlier that in the case of exponential bounce the perturbation modes lie outside of the 
Hubble radius at the distant past, unlike to the case of quasi-matter bounce scenario where the perturbation modes generate far before the bounce 
and thus the modes remain within the deep sub-Hubble regime at the distant past. 
This resolves the horizon problem in the quasi-matter bounce scenario, while the exponential bounce seems to suffer from such problem. Thus in 
the following, we will concentrate on the quasi-matter bounce scenario and perform the perturbation analysis in order to determine the observable quantities 
like the spectral tilt for the curvature perturbation and the tensor-to-scalar ratio respectively. This is the subject of 
Sec.[\ref{sec-perturbation}].\\

In regard to the perturbation analysis, we represent the present entropic cosmology with the ghost free Gauss-Bonnet (GB) theory of gravity 
proposed in \cite{Nojiri:2018ouv}. The motivation of such representation is due to the rich structure of the Gauss-Bonnet theory in various 
directions of cosmology \cite{Odintsov:2022unp,Elizalde:2020zcb,Bamba:2020qdj,Nojiri:2022xdo}.
We briefly demonstrate the essential features of the GB theory, and then we will show that for a 
certain $\gamma(N)$ in the context of entropic cosmology, there exists an equivalent set of GB parameters in the side of Gauss-Bonnet cosmology that 
results to the same cosmological evolution as of the generalized entropy. The 
action for $f(\mathcal{G})$ gravity is given by \cite{Nojiri:2018ouv},
\begin{equation}
\label{FRGBg19} S=\int d^4x\sqrt{-g} \left(\frac{1}{2\kappa^2}R 
+ \lambda \left( \frac{1}{2} \partial_\mu \chi \partial^\mu \chi 
+ \frac{\mu^4}{2} \right) - \frac{1}{2} \partial_\mu \chi \partial^\mu \chi
+ h\left( \chi \right) \mathcal{G} - V\left( \chi \right)\right)\, ,
\end{equation}
where $\mu$ is a constant having mass dimension $[+1]$, $\lambda$ represents the Lagrange multiplier, $\chi$ is a scalar field and $V(\chi)$ is its 
potential. Moreover $\mathcal{G} = R^2 - 4R_{\mu\nu}R^{\mu\nu} + R_{\mu\nu\alpha\beta}R^{\mu\nu\alpha\beta}$ is the Gauss-Bonnet scalar and 
$h(\chi)$ symbolizes the Gauss-Bonnet coupling with the scalar field. The above action results to a ghost free action, as shown in \cite{Nojiri:2018ouv}. 
Varying the Lagrange multiplier, i.e $\frac{\delta S}{\delta \lambda} = 0$ gives the following constraint equation,
\begin{equation}
\label{FRGBg20} 
0=\frac{1}{2} \partial_\mu \chi \partial^\mu \chi + \frac{\mu^4}{2} \, 
\end{equation}
which clearly indicates that the kinetic term of $\chi$ is a constant, thus it may be absorbed within the $V(\chi)$. Therefore the new potential 
of the $\chi$ comes as,
\begin{equation}
\label{FRGBg21} 
\tilde V \left(\chi\right) \equiv \frac{1}{2}
\partial_\mu \chi \partial^\mu \chi + V \left( \chi \right)
= - \frac{\mu^4}{2} + V \left( \chi \right) \, ,
\end{equation}
owing to which, the action of Eq.~(\ref{FRGBg19}) is equivalently re-written as,
\begin{equation}
\label{FRGBg22} 
S=\int d^4x\sqrt{-g} \left(\frac{1}{2\kappa^2}R 
+ \lambda \left( \frac{1}{2} \partial_\mu \chi \partial^\mu \chi 
+ \frac{\mu^4}{2} \right) + h\left( \chi \right) \mathcal{G}
 - \tilde V\left( \chi \right)\right) \, .
\end{equation}
For the above action (\ref{FRGBg22}), the scalar and the gravitational equations have the following form,
\begin{align}
\label{FRGBg23} 
0 =& - \frac{1}{\sqrt{-g}} \partial_\mu \left(
\lambda g^{\mu\nu}\sqrt{-g} \partial_\nu \chi \right)
+ h'\left( \chi \right) \mathcal{G} - {\tilde V}'\left( \chi \right) \, , \\
\label{FRGBg24} 
0 =& \frac{1}{2\kappa^2}\left(- R_{\mu\nu} 
+ \frac{1}{2}g_{\mu\nu} R\right) - \frac{1}{2} \lambda \partial_\mu \chi \partial_\nu \chi
 - \frac{1}{2}g_{\mu\nu} \tilde V \left( \chi \right)
+ D_{\mu\nu}^{\ \ \tau\eta} \nabla_\tau \nabla_\eta h \left( \chi \right)\, ,
\end{align}
where $D_{\mu\nu}^{\ \ \tau\eta}$ is given by,
\begin{align}
D_{\mu\nu}^{\ \ \tau\eta}=&\left( \delta_{\mu}^{\ \tau}\delta_{\nu}^{\ \eta} + \delta_{\nu}^{\ \tau}\delta_{\mu}^{\ \eta} 
 - 2g_{\mu\nu}g^{\tau\eta} \right) R + \left( -4g^{\rho\tau}\delta_{\mu}^{\ \eta}\delta_{\nu}^{\ \sigma}
 - 4g^{\rho\tau}\delta_{\nu}^{\ \eta}\delta_{\mu}^{\ \sigma} + 4g_{\mu\nu}g^{\rho\tau}g^{\sigma\nu} \right) R_{\rho\sigma}\nonumber\\
&+4R_{\mu\nu}g^{\tau\eta} - 2R_{\rho\mu\sigma\nu} \left(g^{\rho\tau}g^{\sigma\nu} + g^{\rho\eta}g^{\sigma\tau}\right)
\nonumber
\end{align}
with having in mind $g^{\mu\nu}D_{\mu\nu}^{\ \ \tau\eta} = 4\left[-\frac{1}{2}g^{\tau\eta}R + R^{\tau\eta} \right]$. Due to the 
FRW metric shown in Eq.(\ref{metric}), and assuming that $\chi$ and $\lambda$ are functions of $t$ only, 
Eq.~(\ref{FRGBg20}) immediately gives the solution for $\chi$ as,
\begin{equation}
\label{frgdS4} \chi = \mu^2 t \, .
\end{equation}
Consequently the temporal and spatial components of Eq.~(\ref{FRGBg24}) become,
\begin{align}
\label{FRGFRW1} 
0 = & - \frac{3H^2}{2\kappa^2}
 - \frac{\mu^4 \lambda}{2} + \frac{1}{2} \tilde V \left( \mu^2 t \right)
 - 12 \mu^2 H^3 h' \left( \mu^2 t \right) \, , \\
\label{FRGFRW2} 
0 = & \frac{1}{2\kappa^2} \left( 2 \dot H + 3 H^2 \right)
 - \frac{1}{2} \tilde V \left( \mu^2 t \right)
+ 4 \mu^4 H^2 h'' \left( \mu^2 t \right) + 8 \mu^2 \left( \dot H +
H^2 \right) H h' \left( \mu^2 t \right) \, ,
\end{align}
and moreover, the scalar field equation comes as,
\begin{equation}
\label{FRGFRW3} 
0 = \mu^2 \dot\lambda + 3 \mu^2 H \lambda + 24 H^2
\left( \dot H + H^2 \right) h'\left( \mu^2 t \right)
 - {\tilde V}'\left( \mu^2 t \right) \, .
\end{equation}
Here it may be mentioned that the above field equations are not independent, in particular, Eq.(\ref{FRGFRW2}) can be achieved from the other two. 
Eq.(\ref{FRGFRW1}) and Eq.(\ref{FRGFRW2}) provide the energy density ($\rho_\mathrm{GB}$) and the pressure ($p_\mathrm{GB}$) 
corresponding to this ghost free Gauss-Bonnet gravity theory as:
\begin{eqnarray}
 \rho_\mathrm{GB}&=&-\mu^4\lambda + \tilde{V}(\chi) - 24\mu^2H^3h'(\chi)~~,\nonumber\\
 p_\mathrm{GB}&=&-\tilde{V}(\chi) + 8\mu^4H^2h''(\chi) + 16\mu^2\left(\dot{H} + H^2\right)Hh'(\chi)~~,
 \label{energy density and pressure-GB}
\end{eqnarray}
respectively. For the Gauss-Bonnet theory, the speed of the gravitational wave ($c_T^2$) is, in general, different than unity and the deviation of $c_T^2$ from unity is 
controlled by the GB coupling function. In particular, the $c_T^2$ in the Gauss-Bonnet theory is,
\begin{equation}
c_T^2 = 1 + \frac{16\left( \ddot{h}-\dot{h}H \right)}{\frac{1}{\kappa^2} + 16\dot{h}H}~~.
\label{gravitaional wave speed}
\end{equation}
Such deviation of $c_T^2$ from unity is not consistent with the GW170817 event which argues that 
the gravitational wave propagates with same speed of light. 
Thus in order to be consistent with the GW170817 event, we consider such class of GB coupling functions that satisfy the following condition 
\cite{Elizalde:2020zcb,Nojiri:2022xdo,Odintsov:2020sqy},
\begin{eqnarray}
 \ddot{h} = \dot{h}H~~~~~~~~~\Longrightarrow~~~~~~~~~~\dot{h} = h_0a(t)~~,
 \label{condition}
\end{eqnarray}
where $h_0$ being the integration constant. 
With the above condition, $\rho_\mathrm{GB}$ and $p_\mathrm{GB}$ from Eq.(\ref{energy density and pressure-GB}) turn out to be,
\begin{eqnarray}
 \rho_\mathrm{GB}&=&-\mu^4\lambda + \tilde{V}(\chi) - 24\mu^2H^3h'(\chi)~~,\nonumber\\
 p_\mathrm{GB}&=&-\tilde{V}(\chi) + 8\mu^2\left(2\dot{H} + 3H^2\right)Hh'(\chi)~~,
 \label{energy density and pressure-GB}
\end{eqnarray}
respectively. 

To represent the entropic cosmology corresponding to the $S_\mathrm{g}[\alpha_{\pm},\beta,\gamma,\epsilon]$ as a ghost free Gauss-Bonnet 
gravity theory compatible with the GW170817 event, we need to compare the above forms of energy density and pressure of the Gauss-Bonnet theory 
with that of coming from the entropy function as obtained in Eq.(\ref{energy density and pressure}), i.e 
$\rho_\mathrm{GB} \equiv \rho_\mathrm{eff}$ and $p_\mathrm{GB} \equiv p_\mathrm{eff}$ respectively. By doing so, we obtain --
\begin{eqnarray}
 \mu^4\lambda - \tilde{V}(\chi) + 24\mu^2H^3h'(\chi) =
 \left(\frac{3\epsilon\alpha}{4\beta G^2}\right)
 \left[\ln{\left\{2\left(\frac{2}{\gamma(N) + \sqrt{\gamma^2(N) + 4}}\right)^{1/\beta} - 1\right\}}\right]^{-1}
\label{compare-1} 
\end{eqnarray}
and
\begin{eqnarray}
 &-&\mu^4\lambda + 16\mu^2\dot{H}Hh'(\chi)\nonumber\\
 &=&H^4\left(\frac{\gamma'(N)}{8\pi^2\gamma(N)}\right)
 \left[\frac{\left\{1 + \frac{1}{\epsilon}\tanh\left(\frac{\epsilon\pi\alpha}{\beta GH^2}\right)\right\}^{\beta} - 
\left\{1 + \frac{1}{\epsilon}\tanh\left(\frac{\epsilon\pi\alpha}{\beta GH^2}\right)\right\}^{-\beta}}
{\alpha~\mathrm{sech}^2\left(\frac{\epsilon \pi\alpha}{\beta GH^2}\right)\left[
\left\{1 + \frac{1}{\epsilon}\tanh\left(\frac{\epsilon \pi\alpha}{\beta GH^2}\right)\right\}^{\beta-1} 
+ \left\{1 + \frac{1}{\epsilon}\tanh\left(\frac{\epsilon \pi\alpha}{\beta GH^2}\right)\right\}^{-\beta-1}\right]}\right]~.\nonumber\\
\label{compare-2}
\end{eqnarray}
The above two algebraic equations can be solved for $\tilde{V}(\chi)$ and $\lambda(t)$ to get,
\begin{eqnarray}
\tilde{V}(\chi)&=&-8\pi G~F_1\left[\gamma(N),\gamma'(N)\right]\left(\frac{1}{\kappa^2} + 8h_0a(t)H(t)\right) \bigg|_{t=\chi/\mu^2}\, ,
\label{equiv potential}\\
\mu^4\lambda(t)&=&-8\pi G~F_2\left[\gamma(N),\gamma'(N)\right]\left(\frac{1}{\kappa^2} - 8h_0a(t)H(t)\right)\, ,
\label{equiv potential and LM}
\end{eqnarray}
where the functions $F_1\left[\gamma(N),\gamma'(N)\right]$ and $F_2\left[\gamma(N),\gamma'(N)\right]$ are given by,
\begin{eqnarray}
 F_1\left[\gamma(N),\gamma'(N)\right]&=&\left(\frac{3\epsilon\alpha}{4\beta G^2}\right)
 \left[\ln{\left\{2\left(\frac{2}{\gamma(N) + \sqrt{\gamma^2(N) + 4}}\right)^{1/\beta} - 1\right\}}\right]^{-1} +
 H^4\left(\frac{\gamma'(N)}{8\pi^2\gamma(N)}\right)\times\nonumber\\
 &\Bigg[&\frac{\left\{1 + \frac{1}{\epsilon}\tanh\left(\frac{\epsilon\pi\alpha}{\beta GH^2}\right)\right\}^{\beta} - 
\left\{1 + \frac{1}{\epsilon}\tanh\left(\frac{\epsilon\pi\alpha}{\beta GH^2}\right)\right\}^{-\beta}}
{\alpha~\mathrm{sech}^2\left(\frac{\epsilon \pi\alpha}{\beta GH^2}\right)\left[
\left\{1 + \frac{1}{\epsilon}\tanh\left(\frac{\epsilon \pi\alpha}{\beta GH^2}\right)\right\}^{\beta-1} 
+ \left\{1 + \frac{1}{\epsilon}\tanh\left(\frac{\epsilon \pi\alpha}{\beta GH^2}\right)\right\}^{-\beta-1}\right]}\Bigg]\nonumber\\
\nonumber
 \end{eqnarray}
 and
 \begin{eqnarray}
 F_2\left[\gamma(N),\gamma'(N)\right] = H^4\left(\frac{\gamma'(N)}{8\pi^2\gamma(N)}\right)
 \left[\frac{\left\{1 + \frac{1}{\epsilon}\tanh\left(\frac{\epsilon\pi\alpha}{\beta GH^2}\right)\right\}^{\beta} - 
\left\{1 + \frac{1}{\epsilon}\tanh\left(\frac{\epsilon\pi\alpha}{\beta GH^2}\right)\right\}^{-\beta}}
{\alpha~\mathrm{sech}^2\left(\frac{\epsilon \pi\alpha}{\beta GH^2}\right)\left[
\left\{1 + \frac{1}{\epsilon}\tanh\left(\frac{\epsilon \pi\alpha}{\beta GH^2}\right)\right\}^{\beta-1} 
+ \left\{1 + \frac{1}{\epsilon}\tanh\left(\frac{\epsilon \pi\alpha}{\beta GH^2}\right)\right\}^{-\beta-1}\right]}\right]\nonumber
\end{eqnarray}
respectively. Eq.(\ref{equiv potential}) and Eq.(\ref{equiv potential and LM}) clearly 
depict that for a certain form of $\gamma(N)$ in the context of entropic cosmology, there exists 
an equivalent $\tilde{V}(\chi)$ and $\lambda$ in the side of Gauss-Bonnet cosmology that results to the same cosmological evolution as of the generalized 
entropy. therefore we may argue that the entropic cosmology of $S_\mathrm{g}$ can be equivalently 
represented by Gauss-Bonnet cosmology where the $\tilde{V}(\chi)$ and $\lambda(t)$ are given by Eq.(\ref{equiv potential}) and 
Eq.(\ref{equiv potential and LM}) respectively.

As mentioned earlier that we will concentrate on the quasi-matter bounce to analyze the perturbation (see the next subsection), in which case, 
the $\gamma(N)$ is shown in Eq.(\ref{matter bounce-6}), and thus, 
$F_1\left[\gamma(N),\gamma'(N)\right]$ and $F_2\left[\gamma(N),\gamma'(N)\right]$ come as,
\begin{eqnarray}
 F_1\left[\gamma(N),\gamma'(N)\right]&=&\left(\frac{\epsilon\pi\sqrt{\alpha}}{8\pi\beta G^2}\right)e^{-2N/\sqrt{\alpha}}
 \left(e^{N/\sqrt{\alpha}} - 2\right)\nonumber\\
 F_2\left[\gamma(N),\gamma'(N)\right]&=&\left(\frac{\epsilon\pi\alpha}{8\pi\beta G^2}\right)e^{-2N/\sqrt{\alpha}}
 \left\{3\left(e^{N/\sqrt{\alpha}} - 1\right) + \frac{1}{\sqrt{\alpha}}\left(e^{N/\sqrt{\alpha}} - 2\right)\right\}~~.
 \label{equiv-MBS}
\end{eqnarray} 
Thus as a whole -- Eq.(\ref{equiv potential}) and Eq.(\ref{equiv potential and LM}) establish the equivalence 
between the entropic cosmology corresponding to the $S_\mathrm{g}$ and the Gauss-Bonnet cosmology, and moreover, Eq.(\ref{equiv-MBS}) 
shows such equivalence in the case of quasi-matter bounce scenario.

\subsection*{Cosmological perturbation and phenomenology of the entropic quasi-matter bounce}\label{sec-perturbation}
As mentioned earlier that we consider the quasi-matter bounce scenario described by the scale factor (\ref{matter bounce-1}) to analyze the perturbation, where 
the perturbation modes generate during the contracting phase deep in the sub-Hubble regime, which in turn ensures the resolution of 
the horizon problem. For the scale factor (\ref{matter bounce-1}), the Ricci scalar during the contracting era is given by,
\begin{eqnarray}
 R(t) = \frac{12n(1-4n)}{t^2}~~,
 \label{useful quantities}
\end{eqnarray}
Due to this expression of $R = R(t)$, one can achieve the scale factor, the Hubble parameter and its derivative (during the contracting era) 
in terms of the Ricci scalar as follows:
\begin{eqnarray}
a(R)&=&\frac{a_0^{n}}{\left(\widetilde{R}/R_0\right)^{n}}~~~~,~~~~
H(R) = -2n\widetilde{R}^{1/2}~~~~~\mathrm{and}~~~~~\dot{H}(R)=-2n\widetilde{R}~~,
\label{quantities-deep-contracting-1}
\end{eqnarray}
where $R_0 = \frac{1}{t_0^2}$ and $\widetilde{R}(t) = \frac{R(t)}{12n(1-4n)}$. It may be noted that $n$, $a_0$ and $t_0$ are related 
to the entropic parameters via Eq.(\ref{matter bounce-2}). Moreover from Eq.(\ref{condition}), we determine the 
derivative of the GB coupling function (in terms of the Ricci scalar) as
\begin{eqnarray}
 \dot{h}(R) = \frac{(2n+1)}{8\pi G\sqrt{R_0}}\left(\frac{R_0}{\widetilde{R}}\right)^n~~,
 \label{quantities-deep-contracting-2}
\end{eqnarray}
where the integration constant $h_0$ is adjusted in suitable manner. 
With the above expressions of $H(R)$ and $\dot{h}(R)$, the functions $Q_i$ in the context of 
the ghost free Gauss-Bonnet theory of gravity \cite{Hwang:2005hb,Noh:2001ia} come with the following expressions,
\begin{align}
Q_a=&-8\dot{h}H^2 = -\frac{4n^2(1+2n)\sqrt{\widetilde{R}}}{\pi G}\left(\frac{\widetilde{R}}{R_0}\right)^{\frac{1}{2}-n}\, , \nonumber\\
Q_b=&-16\dot{h}H = \frac{4n(1+2n)}{\pi G}\left(\frac{\widetilde{R}}{R_0}\right)^{\frac{1}{2}-n}\, ,\nonumber\\
Q_c=&Q_d = 0 \, ,\nonumber\\
Q_e=&-32\dot{h}\dot{H} = \frac{8n(1+2n)\sqrt{\widetilde{R}}}{\pi G}\left(\frac{\widetilde{R}}{R_0}\right)^{\frac{1}{2}-n}\, ,\nonumber\\
Q_f=&16 \left[ \ddot{h} - \dot{h}H \right] = 0 \, ,
\label{Q-s}
\end{align}
respectively. The last expression of Eq.(\ref{Q-s}), i.e. $Q_f = 0$, is a direct consequence of the fact that the speed of the gravitational 
wave is unity. Finally Eq.~(\ref{equiv potential and LM}) gives the Lagrange multiplier function as,
\begin{eqnarray}
\mu^4\lambda = -\frac{n\widetilde{R}}{2\pi G}\left[1 - 16n(1+2n)\left(\frac{\widetilde{R}(t)}{R_0}\right)^{\frac{1}{2} - n}\right]\, .
\label{LM-deep-contracting}
\end{eqnarray}
We will frequently use these expressions in the subsequent sections. 

\subsubsection*{Scalar perturbation}
In the comoving gauge, the second order action for primordial curvature perturbation (symbolized by $\Psi(t,\vec{x})$) comes as 
\cite{Hwang:2005hb,Noh:2001ia},
\begin{align}
\delta S_{\psi} = \int dt d^3\vec{x} a(t) z(t)^2\left[\dot{\Psi}^2
 - \frac{c_s^2}{a^2}\left(\partial_i\Psi\right)^2\right]\, .
\label{sp2}
\end{align}
We have shown that the energy density and the pressure corresponds to the entropy $S_\mathrm{g}$, i.e $\rho_\mathrm{g}$ and $p_\mathrm{g}$, 
can be represented by a ghost free $f(R,\mathcal{G})$ gravity theory for suitable forms of scalar field potential and the GB coupling function. 
As a result, $z(t)$ and $c_s^2$ have the following forms \cite{Hwang:2005hb},
\begin{align}
z(t) = \frac{a(t)}{H + \frac{Q_a}{2F + Q_b}} \sqrt{-\mu^4\lambda + \frac{3Q_a^2 + Q_aQ_e}{2F + Q_b}}
\label{sp3}
\end{align}
and
\begin{align}
c_{s}^{2} = 1 + \frac{Q_aQ_e/\left(2F + Q_b\right)}{-\mu^4\lambda + 3\frac{Q_a^2}{2F + Q_b}} \, ,
\label{speed scalar perturbation}
\end{align}
respectively, with and $F=\frac{1}{16\pi G}$. Plugging the expressions of $Q_i$ into Eq.(\ref{sp3}) yields the form of $z(t)$ as,
\begin{align}
z(t) = -\frac{a_0^n}{\kappa \left(\widetilde{R}/R_0\right)^{n}}~\frac{\sqrt{P(R)}}{Q(R)}
\label{sp4}
\end{align}
where $P(R)$ and $Q(R)$ are given by,
\begin{align}
P(R) = 4n\left[1 - 16n(1+2n)\left(\frac{\widetilde{R}}{R_0}\right)^{\frac{1}{2} - n} 
+ \mathcal{O}\left(\frac{\widetilde{R}}{R_0}\right)^{1-2n}\right]\, ,
\label{P}
\end{align}
and
\begin{align}
Q(R) = 2n\left[1 + 16n(1+2n)\left(\frac{\widetilde{R}}{R_0}\right)^{\frac{1}{2} - n} 
+ \mathcal{O}\left(\frac{\widetilde{R}}{R_0}\right)^{1-2n}\right]\, ,
\label{Q}
\end{align}
respectively. Recall that the perturbation modes generate far before the bounce when the Ricci scalar satisfies the condition 
$\frac{\widetilde{R}}{R_0} \ll 1$ due to $\widetilde{R} \rightarrow 0$ at $t \rightarrow -\infty$. This results to 
$z^2(t) > 0$ which in turn ensures the stability of the curvature perturbation.

In order to solve the Mukhanov-Sasaki equation for the curvature perturbation, we will use conformal time defined by $\eta = \int\frac{dt}{a(t)}$ which, 
due to $a(t) \sim t^{2n}$, comes as,
\begin{eqnarray}
\eta(t) = \left[\frac{1}{a_0^n(1-2n)}\right]t^{1-2n}\, .
\label{conformal time}
\end{eqnarray}
Clearly $\eta(t)$ is a monotonic increasing function of $t$ (recall that $n < 1/2$). With the above expression of $\eta=\eta(t)$, the Ricci scalar 
can be obtained in terms of conformal time as follows:
\begin{eqnarray}
\widetilde{R}(\eta) = \frac{1}{\left[a_0^n(1-2n)\right]^{2/(1-2n)}}\times\frac{1}{\eta^{2/(1-2n)}} \propto \frac{1}{\eta^{2/(1-2n)}}\,.
\label{ricci-scalar-conformal-time}
\end{eqnarray}
by using which, the $z(\eta)$ from Eq.(\ref{sp4}) turns out to be,
\begin{eqnarray}
z(\eta) \propto \left(\frac{\sqrt{P(\eta)}}{Q(\eta)}\right)\eta^{2n/(1-2n)}\, ,
\label{z-eta}
\end{eqnarray}
with $P(\eta) = P(R(\eta))$ and $Q(\eta) = Q(R(\eta))$. 
Consequently the factor $\frac{1}{z}\frac{d^2z}{d\eta^2}$ (which demonstrates the interaction of the curvature perturbation with the background evolution 
in the Mukhanov-Sasaki equation) is given by,
\begin{eqnarray}
\frac{1}{z}\frac{d^2z}{d\eta^2} = \frac{\xi(\xi - 1)}{\eta^2}\left\{1 + 24\left(1-4n^2\right)\left(\frac{\widetilde{R}}{R_0}\right)^{\frac{1}{2} - n} 
+ \mathcal{O}\left(\frac{\widetilde{R}}{R_0}\right)^{1-2n}\right\}
\label{derivative-z-eta}
\end{eqnarray}
where we use $\frac{d\widetilde{R}}{d\eta} = \frac{-2}{(1-2n)}\frac{\widetilde{R}}{\eta}$ and moreover $\xi = \frac{2n}{(1-2n)}$ in the above expression. 
Furthermore the speed of the scalar perturbation from Eq.~(\ref{speed scalar perturbation}) comes with the following expression,
\begin{eqnarray}
c_s^2 = 1 + \mathcal{O}\left(\frac{\widetilde{R}}{R_0}\right)^{1-2n}\, .
\label{sound-speed}
\end{eqnarray}
Having all the necessary ingredients in hand, we now introduce the canonical Mukhanov-Sasaki (MS) 
variable: $v(\eta,\vec{x}) = z(\eta)\Psi(\eta,\vec{x})$, and consequently the Fourier version of Mukhanov-Sasaki equation is, 
\begin{eqnarray}
\frac{d^2v_k(\eta)}{d\eta^2} + \left(c_s^2k^2 - \frac{1}{z}\frac{d^2z}{d\eta^2}\right)v_k(\eta) = 0\, ,
\label{scalar-MS-equation}
\end{eqnarray}
with $v_k(\eta)$ being the Fourier mode for $v(\eta,\vec{x})$. 
Eq.(\ref{scalar-MS-equation}) clearly depicts that the dynamics of $v_k(\eta)$ is controlled by the background quantities like 
$z''/z$ and $c_s^2$. As demonstrated after Eq.~(\ref{Q})) that the Ricci scalar 
satisfies $\frac{\widetilde{R}}{R_0} \ll 1$ during the contracting era, and thus one can retain the leading order term of 
$\left(\widetilde{R}/R_0\right)^{\frac{1}{2} - n}$ in the expressions of $z''/z$ and $c_s^2$. In effect, they take the following forms,
\begin{eqnarray}
\frac{1}{z}\frac{d^2z}{d\eta^2}&=&\frac{\xi(\xi - 1)}{\eta^2}\left[1 + 24\left(1-4n^2\right)\left(\frac{\widetilde{R}}{R_0}\right)^{\frac{1}{2} - n}
\right]~~,\nonumber\\
c_s^2&=&1\, .
\label{approximate-behaviour}
\end{eqnarray}
Moreover due to $n < 1/2$ (in order to resolve the horizon problem) along with $\frac{\widetilde{R}}{R_0} \ll 1$ reveal that 
the term $\left(\widetilde{R}/R_0\right)^{\frac{1}{2} - n}$ within the parenthesis can be safely considered to be small during the contracting era when the 
perturbation modes cross the horizon. In effect, $z''/z$ becomes proportional to $1/\eta^2$, in particular, $\frac{1}{z}\frac{d^2z}{d\eta^2} = \sigma/\eta^2$, where,
\begin{eqnarray}
\sigma = \xi(\xi - 1)\left[1 + 24\left(1-4n^2\right)\left(\frac{\widetilde{R}}{R_0}\right)^{\frac{1}{2} - n}\right]\, ,
\label{sigma}
\end{eqnarray}
which is approximately a constant during the generation era of the perturbation modes in the sub-Hubble regime. Using 
$z''/z \propto \eta^{-2}$ and $c_s^2 = 1$, one may solve $v_k(\eta)$ from Eq.~(\ref{scalar-MS-equation}) and is given by,
\begin{eqnarray}
v(k,\eta) = \frac{\sqrt{\pi|\eta|}}{2}\left[c_1(k)H_{\nu}^{(1)}(k|\eta|) + c_2(k)H_{\nu}^{(2)}(k|\eta|)\right]\, ,
\label{scalar-MS-solution}
\end{eqnarray}
with $\nu = \sqrt{\sigma + \frac{1}{4}}$, and, 
$H_{\nu}^{(1)}(k|\eta|)$ and $H_{\nu}^{(2)}(k|\eta|)$ symbolize the Hermite functions (having order 
$\nu$) of first and second kind, respectively. The integration constants $c_1$ and $c_2$ can be determined from the Bunch-Davies condition of the MS variable. 
The Bunch-Davies vacuum of the MS variable during $\eta \rightarrow -\infty$, i.e $\lim_{k|\eta| \gg 1}v(k,\eta) = \frac{1}{\sqrt{2k}}e^{-ik\eta}$, 
is well justified due to the fact that the perturbation modes lie within the sub-Hubble radius at the distant past. This results to 
$c_1 = 0$ and $c_2 = 1$ respectively. Owing to which, the scalar power spectrum for $k$th mode becomes (defined by 
$\mathcal{P}_{\Psi}(k,\eta) = \frac{k^3}{2\pi^2} \left| \frac{v(k,\eta)}{z(\eta)} \right|^2$),
\begin{eqnarray}
\mathcal{P}_{\Psi}(k,\eta) = \frac{k^3}{2\pi^2} \left| \frac{\sqrt{\pi|\eta|}}{2z (\eta)}H_{\nu}^{(2)}(k|\eta|) \right|^2\, ,
\label{scalar-power-spectrum}
\end{eqnarray}
where the solution of $v(k,\eta)$ is used. The horizon crossing condition for $k$th mode is given by 
$k = |aH|$ which, due to Eq.~(\ref{quantities-deep-contracting}), is determined as,
\begin{eqnarray}
k = \frac{1}{\left| \eta_h\right|}\left(\frac{2n}{1-2n}\right) \quad \Rightarrow \quad k\left| \eta_h\right| = \frac{2n}{1-2n}\, ,
\label{hc-1}
\end{eqnarray}
with the suffix 'h' symbolizes the instant of horizon crossing. The observable quantities like the spectral tilt for the curvature perturbation and the tensor to 
scalar ratio are eventually determined around the large scale modes given by $k = 0.05\mathrm{Mpc}^{-1}$. Thus Eq.(\ref{hc-1}) leads to the horizon crossing 
instant for $k = 0.05\mathrm{Mpc}^{-1}$ as $\eta_h \approx -13\,\mathrm{By}$. This is however expected because of the following reasons: 
(1) the universe's evolution is symmetric around the bounce for the scale factor in Eq.(\ref{matter bounce-1}), and (2) the large scale modes re-enter 
the horizon near the present epoch of the universe $\sim 13.5\mathrm{By}$. This immediately leads to the estimation of the Ricci scalar at the horizon 
crossing of the large scale modes as $\frac{\widetilde{R}}{R_0} \sim 10^{-6}$ (with $n = 0.3$, $R_0 = 1\mathrm{By}^{-2}$ and 
$a_0 \sim \mathcal{O}(1)$ which will be shown to be consistent with the viability of the model in respect to the Planck data, see the next subsection). 
This in turn justifies the condition $\frac{\widetilde{R}}{R_0} \ll 1$ considered earlier in determining $z(t)$.

In order to determine the spectral tilt, we need the scalar power spectrum in the super-Hubble regime when the mode satisfies 
$k|\eta| < 2n/(1-2n)$, and consequently, $\mathcal{P}_{\Psi}(k,\eta)$ becomes,
\begin{eqnarray}
\mathcal{P}_{\Psi}(k,\eta) = \left[\left(\frac{1}{2\pi}\right)\frac{1}{z\left|\eta\right|}\frac{\Gamma(\nu)}{\Gamma(3/2)}\right]^2
\left(\frac{k|\eta|}{2}\right)^{3-2\nu}\, ,
\label{scalar-power-spectrum-superhorizon}
\end{eqnarray}
recall that $\nu = \sqrt{\sigma + \frac{1}{4}}$. The above expression of $\mathcal{P}_{\Psi}(k,\eta)$ leads to the spectral tilt for the curvature 
perturbation, symbolized by $n_s$. However before calculating the $n_s$, we perform the tensor perturbation which is required for the other observable 
quantity namely the tensor-to-scalar ratio ($r$).

\subsubsection*{Tensor perturbation}
The quadratic order action of the tensor perturbation is \cite{Hwang:2005hb,Noh:2001ia},
\begin{align}
\delta S_{h} = \int dt d^3\vec{x} a(t) z_T(t)^2\left[\dot{h}_{ij}\dot{h}^{ij}
 - \frac{1}{a^2}\left(\partial_kh_{ij}\right)^2\right]\, .
\label{tp2}
\end{align}
As demonstrated earlier that the entropic energy density and the entropic pressure corresponding to the $S_\mathrm{g}$ are represented 
by Lagrange multiplier $f(R,\mathcal{G})$ theory, in which case, the function $z_T$ takes the following form \cite{Hwang:2005hb},
\begin{align}
z_T(t) = a\sqrt{F + \frac{1}{2}Q_b}\, ,
\label{tp3}
\end{align}
where $F = \frac{1}{16\pi G}$ and the $Q_b$ is given in Eq.~(\ref{Q-s}). It may be observed from Eq.(\ref{tp2}) that the speed of the gravitational wave 
is unity, as the GB coupling function in the present context satisfies $\ddot{h} = \dot{h}H$ which leads to $c_T^2 = 1$ and ensures the compatibility 
of the bounce model with the GW170817 event. We use Eq.~(\ref{quantities-deep-contracting-1}) to get $z_T$ as follows, 
\begin{align}
z_T=\frac{a_0^n}{\sqrt{2}\kappa\widetilde{R}^n}\left[1 + 16n(1+2n)\left(\frac{\widetilde{R}}{R_0}\right)^{\frac{1}{2} - n}\right]\, ,
\label{zT-1}
\end{align}
which depicts that $z_T^2 > 0$ and thus the tensor perturbation is stable. 
Eq.~(\ref{ricci-scalar-conformal-time}) immediately leads to $z_T$ in terms of the conformal time as,
\begin{eqnarray}
z_T(\eta) \propto S(R(\eta))\eta^{2n/(1-2n)}\, ,
\label{zT-2}
\end{eqnarray}
where $S(R(\eta))$ has the following form,
\begin{eqnarray}
S(R(\eta)) = 1 + 16n(1+2n)\left(\frac{\widetilde{R}}{R_0}\right)^{\frac{1}{2} - n}\, .
\label{S}
\end{eqnarray}
Consequently we determine the factor $z_T''/z_T$:
\begin{eqnarray}
\frac{1}{z_T}\frac{d^2z_T}{d\eta^2} = \frac{\xi(\xi-1)}{\eta^2}\left[1 - 16(1-4n^2)\left(\frac{\widetilde{R}}{R_0}\right)^{\frac{1}{2} - n}
\right]\, .
\label{derivative-zT}
\end{eqnarray}
The above expression will be useful for solving the tensor Mukhanov-Sasaki equation. 
The condition $\frac{\widetilde{R}}{R_0} \ll 1$ together with $n < 1/2$ clearly demonstrate that the term containing 
$\left(\frac{\widetilde{R}}{R_0}\right)^{\frac{1}{2} - n}$ within the parenthesis can be safely considered to be small and slowly varying 
during the contracting phase. In effect of which, $z_T''/z_T$ becomes proportional to $1/\eta^2$, in particular, 
$\frac{1}{z_T}\frac{d^2z_T}{d\eta^2} = \sigma_T/\eta^2$, with
\begin{eqnarray}
\sigma_T = \xi(\xi - 1)\left[1 - 16(1-4n^2)\left(\frac{\widetilde{R}}{R_0}\right)^{\frac{1}{2} - n}\right]\, .
\label{sigma-T}
\end{eqnarray}
Thus the tensor Mukhanov-Sasaki (MS) equation turns out to be,
\begin{align}
\frac{d^2v_T(k,\eta)}{d\eta^2} + \left(k^2 - \frac{\sigma_T}{\eta^2}\right)v_T(k,\eta) = 0\, ,
\label{tensor-MS-equation}
\end{align}
with $v_T(k,\eta)$ is the Fourier transformed quantity of the tensor MS variable that is defined by $\left(v_T\right)_{ij} = z_Th_{ij}$. 
Here it may be mentioned that both the tensor polarization modes ($+$ and $\times$ polarization modes) obey the same evolution Eq.(\ref{tensor-MS-equation}) -- 
this means that the two polarization modes equally contribute to the energy density of the tensor perturbation variable, and thus 
we will multiply by the factor '2' in the final expression of the tensor power spectrum. Solving Eq.(\ref{tensor-MS-equation}), one gets
\begin{eqnarray}
v_T(k,\eta) = \frac{\sqrt{\pi|\eta|}}{2}\left[D_1~H_{\theta}^{(1)}(k|\eta|) + D_2~H_{\theta}^{(2)}(k|\eta|)\right]\, ,
\label{n1}
\end{eqnarray}
with $\theta = \sqrt{\sigma_T + \frac{1}{4}}$, and, $H_{\theta}^{(1)}(k|\eta|)$ and $H_{\theta}^{(2)}(k|\eta|)$ symbolize the Hermite functions (having order
$\theta$) of first and second kind, respectively. Here it may be mentioned that the solution of $v_T(k,\eta)$ can also be written in terms of cylinder function of order $\theta$. Moreover $D_1$ and $D_2$ are two integration constants that can be determined from the initial condition of $v_T(k,\eta)$. Similar to the curvature perturbation variable, the tensor perturbation initiates from the Bunch-Davies vacuum at the distant past,
i.e. $v_T(k,\eta)$, i.e $\lim_{k|\eta| \gg 1}v_T(k,\eta) = \frac{1}{\sqrt{2k}}e^{-ik\eta}$, which immediately leads to $D_1 = 1$ and $D_2 = 0$. As a result, we obtain the tensor power spectrum for $k$th mode in the super-Hubble regime as,
\begin{align}
\mathcal{P}_{T}(k,\tau) = 2\left[\frac{1}{2\pi}\frac{1}{z_T\left|\eta\right|}\frac{\Gamma(\theta)}{\Gamma(3/2)}\right]^2 \left(\frac{k|\eta|}{2}
\right)^{3 - 2\theta}\, .
\label{tensor-power-spectrum}
\end{align}
where $\theta = \sqrt{\sigma_T + \frac{1}{4}}$.\\

\begin{figure}[!h]
\begin{center}
\centering
\includegraphics[width=3.0in,height=3.0in]{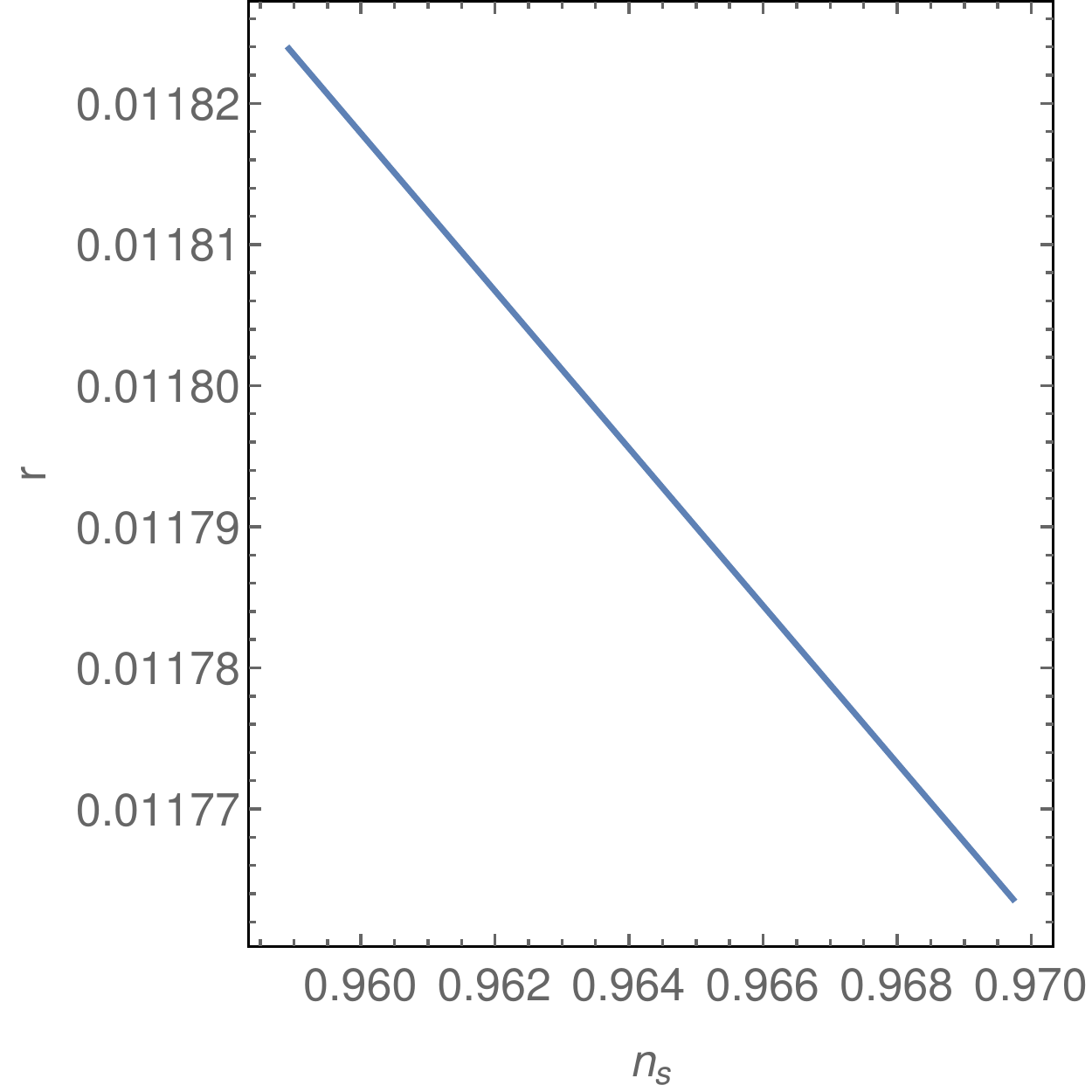}
\caption{Parametric plot of $n_s$ (along $x$-axis) vs. $r$ (along $y$-axis) 
with respect to $n$. Here we take $\alpha = [0.0938,0.0939]$ and $\beta = \frac{\pi}{16}$.}
\label{plot-observable}
\end{center}
\end{figure}

Having obtained the scalar and tensor power spectra, we now evaluate 
the observable quantities like the spectral tilt for the curvature perturbation ($n_s$) and 
the tensor-to-scalar ratio ($r$) respectively, which are defined by,
\begin{eqnarray}
n_s = 1 + \left. \frac{\partial \ln{\mathcal{P}_{\Psi}}}{\partial \ln{k}} \right|_{h} \, , \quad r=\mathcal{P}_T/\mathcal{P}_{\Psi}\, ,
\label{obs-1}
\end{eqnarray}
where the suffix 'h' denotes the horizon crossing of the large scale modes at which we will calculate the $n_s$ and $r$. 
The observational constraints on $n_s$ and $r$ reported by the latest Planck 2018 data \cite{Akrami:2018odb} are,
\begin{eqnarray}
n_s = 0.9649 \pm 0.0042 \quad \mbox{and} \quad r < 0.064 \, .
\label{observable-Planck constraint}
\end{eqnarray}
By using Eq.~(\ref{scalar-power-spectrum-superhorizon}) and 
Eq.~(\ref{tensor-power-spectrum}), we obtain the theoretical expressions of $n_s$ and $r$ as,
\begin{eqnarray}
n_s = 4 - \sqrt{1 + 4\sigma_h} \, , \quad r = 2\left[\frac{z(\eta_h)}{z_T(\eta_h)}\frac{\Gamma(\theta)}{\Gamma(\nu)}\right]^2
\left( k\left|\eta_h\right| \right)^{2(\nu-\theta)}\, ,
\label{obs-2}
\end{eqnarray}
where the quantities have the following forms,
\begin{align}
\nu=&\sqrt{\sigma_h + \frac{1}{4}}\, ; \quad \sigma_h = \xi(\xi - 1)\left[1 + 24\left(1-4n^2\right)\left(\frac{\widetilde{R}_h}{R_0}\right)^{\frac{1}{2} - n}
\right]\, ,\nonumber\\
\theta=&\sqrt{\sigma_{T,h} + \frac{1}{4}} \, ; \quad \sigma_{T,h} = \xi(\xi - 1)\left[1 - 16(1-4n^2)\left(\frac{\widetilde{R}_h}{R_0}\right)^{\frac{1}{2} - n}
\right]\, ,\nonumber\\
z(\eta_h)=&-\frac{1}{\sqrt{n}}\left(\frac{a_0^n}{\kappa\widetilde{R}_h^{n}}\right)\left[1 - 24n(1+2n)
\left(\frac{\widetilde{R}_h}{R_0}\right)^{\frac{1}{2} - n}\right]\, ,\nonumber\\
z_T(\eta_h)=&\frac{1}{\sqrt{2}}\left(\frac{a_0^n}{\kappa\widetilde{R}_h^{n}}\right)\left[1 + 16n(1+2n)
\left(\frac{\widetilde{R}_h}{R_0}\right)^{\frac{1}{2} - n}\right]\, .
\label{obs-3}
\end{align}
Here $\widetilde{R}_h$ represents the Ricci scalar at the horizon crossing, and due to Eq.~(\ref{ricci-scalar-conformal-time}), it turns out to be,
\begin{eqnarray}
\widetilde{R}_h = \left[\frac{1}{a_0^n(1-2n)\left|\eta_h\right|}\right]^{2/(1-2n)}\, ,
\label{obs-4}
\end{eqnarray}
with $\eta_h$ is given by,
\begin{eqnarray}
\left|\eta_h\right| = \left(\frac{2n}{1-2n}\right)\frac{1}{k} \approx \left(\frac{2n}{1-2n}\right)\times13\,\mathrm{By}\, .
\label{obs-5}
\end{eqnarray}
In the second equality of the above equation, we use $k = 0.05\mathrm{Mpc}^{-1} \approx 13\mathrm{By}$ 
(recall that the large scale modes crosses the horizon at around $-13\mathrm{By}$, and being the universe is symmetric, it re-enters 
the horizon during the expanding phase of 
the universe around $+13\mathrm{By}$). 
Using Eq.~(\ref{obs-4}) and Eq.(\ref{obs-5}), one gets $\widetilde{R}_h$ in terms of $n$ and $a_0$:
\begin{eqnarray}
\widetilde{R}_h = \left[\frac{1}{26na_0^n}\right]^{2/(1-2n)}\mathrm{By}^{-2}\, .
\label{obs-6}
\end{eqnarray}
Therefore it is clear that $n_s$ and $r$ in the present context depends on the parameters $n$ and $a_0$. Here we need to recall that $n$ 
and $a_0$ are related to the entropic parameters as $n = \sqrt{\alpha}$ and $a_0 = \pi/\left(4\beta\right)$ respectively. It turns out that the theoretical 
predictions for $n_s$ and $r$ get simultaneously compatible with the recent Planck data for a small range of the entropic parameters 
given by: $\alpha = [0.0938,0.0939]$ and $\beta = \frac{\pi}{16}$ -- this is depicted in Fig.[\ref{plot-observable}]. It may be observed that 
the viable range of $\alpha = [0.0938,0.0939]$ slightly differs than $\alpha = \frac{1}{9}$ which leads to a matter-bounce scenario. Therefore 
in the present context of entropic cosmology, we may argue that a quasi matter bounce scenario gets a consistent $n_s$ as well as a consistent 
$r$ under the Gauss-Bonnet representation. This is unlike to the case when the entropic cosmology under consideration 
is represented by a scalar-tensor representation, in which case, the quasi matter bounce may give a correct $n_s$ 
however the tensor-to-scalar ratio becomes too large to be consistent with the Planck data.  

\section{Conclusion}

We propose a new five-parameter entropy function ($S_\mathrm{g}$) that generalizes the Tsallis, Barrow, R\'{e}nyi, Sharma-Mittal, 
Kaniadakis and Loop Quantum Gravity entropies for suitable limits of the parameters, and at the same time, is also non-singular during the evolution of 
the universe. The non-singular generalized entropy function obeys the generalized third law of thermodynamics, i.e $S_\mathrm{g}$ tends to zero 
for $S \rightarrow 0$ (with $S$ being the Bekenstein-Hawking entropy), and moreover, $S_\mathrm{g}$ turns out to be a monotonic increasing function with 
$S$. Here we would like to mention that beside the five-parameter entropy function, we also proposed a four-parameter entropy function in one 
of our earlier works \cite{Nojiri:2022dkr}, which is able to generalize all the known entropies mentioned above. 
However such an entropy with the four parameters becomes singular (or diverges) when the Hubble parameter vanishes (i.e $H = 0$, for instance, 
in a bounce scenario at the instant of bounce), unlike to the five-parameter entropy function (proposed in the present work) which 
proves to be singular-free during the entire cosmological evolution of the universe. In this regard, we give the following conjecture -- 
``The minimum number of parameters required in a generalized entropy function that can generalize 
all the aforementioned known entropies, and is also singular-free during the universe's evolution -- is equal to five''. 

It is important to note that unlike to the aforementioned known entropies, the newly proposed five-parameter entropy function ($S_\mathrm{g}$) 
is singular-free at $H = 0$. Such $non-singular$ behaviour of the entropy function is useful in describing bouncing
scenario, in which case, the universe undergoes through $H = 0$ at the instant of bounce. Thus we address the cosmological 
implications of $S_\mathrm{g}$ during the early phase of the universe. In particular, 
the effective energy density and the effective pressure sourced from the $S_\mathrm{g}$ modify the Friedmann equations, and 
consequently, we examine whether the cosmology corresponding to $S_\mathrm{g}$ can trigger a non-singular bouncing universe. It turns out that 
being the entropic parameters are constant, the entropic cosmology corresponding to $S_\mathrm{g}$ leads to a constant Hubble parameter as the only possible 
solution. Clearly $H = \mathrm{constant}$ does not lead to the correct evolution of the universe. Thus in order to have an acceptable
cosmological evolution in the present context, we consider the parameters of $S_\mathrm{g}[\alpha_{\pm},\beta,\gamma,\epsilon]$ vary with time, 
in particular, we consider the parameter $\gamma$ to vary with time, and all the other parameters remain fixed, i.e. $\gamma = \gamma(N)$ where 
$N$ represents the e-fold number of the universe. With $\gamma = \gamma(N)$, the Friedmann equations corresponding to $S_\mathrm{g}$ 
depict the following points -- (1) the Hubble parameter is no more a constant during the universe's evolution, and the evolution of $H(N)$ depends 
on the form of $\gamma(N)$, and (2) the Friedmann equations contain a quadratic power of the Hubble parameter, due to which, it allows 
a positive branch as well as a negative branch solution of $H(N)$, which in turn leads to a natural possibility of symmetric bounce 
in the present context of singular-free entropic cosmology. Consequently, we determine the form of $\gamma(N)$ in two different symmetric bounce scenarios, 
in particular, for an exponential bounce and for a quasi-matter bounce respectively. It turns out that similar to the 
entropy function, the functional form of $\gamma(N)$ is of hyperbolic nature in both the bounce scenarios. Despite such similarity, the evolution 
of comoving Hubble radius makes the bounce scenarios qualitatively different from each other. In the case of exponential 
bounce, the comoving Hubble radius asymptotically goes to zero at both sides of the bounce, and thus the perturbation modes generate near the bounce 
when the comoving Hubble radius is infinite in size to contain all the perturbation modes within it. This may result to the ``horizon probelm'' 
which makes the exponential bounce less viable. 
On other hand, the comoving Hubble radius in the case of quasi-matter bounce diverges to infinity at the distant past, owing to which, the perturbation 
modes generate far before the bounce in the deep sub-Hubble regime. As a result, the ``horizon issue'' gets resolved in quasi-matter bounce, as 
the perturbation modes remain in sub-Hubble regime at the distant past. Based on these arguments, we concentrate on the quasi-matter 
bounce scenario and consequently perform the detailed perturbation analysis in order to estimate the observable quantities 
in the present context of entropic cosmology. The observable quantities like the spectral tilt for curvature perturbation ($n_s$) and 
the tensor-to-scalar ratio ($r$) are found to depend on the entropic parameters, as expected. It turns out that the theoretical expectations 
of $n_s$ and $r$ are simultaneously compatible with the recent Planck 2018 data for a small range of the entropic parameters 
given by: $\alpha_+ = \alpha_- = [0.0938,0.0939]$ and $\beta = \frac{\pi}{16}$ respectively.

Furthermore, we show that the entropic cosmology from the proposed 
singular-free entropy is equivalent to holographic cosmology with suitable forms of holographic cut-offs. In particular, the holographic cut-offs 
are determined in terms of either particle horizon and its derivative or future horizon and its derivative.

\section*{Acknowledgments}

This work was supported by MINECO (Spain), project PID2019-104397GB-I00 and  also partially supported by the program Unidad de 
Excelencia Maria de Maeztu CEX2020-001058-M, Spain (SDO). 
This research was also supported in part by the 
International Centre for Theoretical Sciences (ICTS) for the online program - Physics of the Early Universe (code: ICTS/peu2022/1) (TP).


\begin{thebibliography}{99}
 \bibitem{Bekenstein:1973ur}
J.~D.~Bekenstein,
Phys. Rev. D \textbf{7} (1973), 2333-2346
doi:10.1103/PhysRevD.7.2333


\bibitem{Hawking:1975vcx}
S.~W.~Hawking,
Commun. Math. Phys. \textbf{43} (1975), 199-220
[erratum: Commun. Math. Phys. \textbf{46} (1976), 206]
doi:10.1007/BF02345020

\bibitem{Bardeen:1973gs}
J.~M.~Bardeen, B.~Carter and S.~W.~Hawking,
Commun. Math. Phys. \textbf{31} (1973), 161-170
doi:10.1007/BF01645742

\bibitem{Wald:1999vt}
R.~M.~Wald,
Living Rev. Rel. \textbf{4} (2001), 6
doi:10.12942/lrr-2001-6
[arXiv:gr-qc/9912119 [gr-qc]].

\bibitem{Tsallis:1987eu}
C.~Tsallis,
J. Statist. Phys. \textbf{52} (1988), 479-487
doi:10.1007/BF01016429

\bibitem{Renyi}
A.~R\'{e}nyi, Proceedings of the Fourth Berkeley Symposium on Mathematics, Statistics and Probability, University of California Press (1960), 547-56.

\bibitem{Barrow:2020tzx}
J.~D.~Barrow,
Phys. Lett. B \textbf{808} (2020), 135643
doi:10.1016/j.physletb.2020.135643
[arXiv:2004.09444 [gr-qc]].

\bibitem{SayahianJahromi:2018irq}
A.~Sayahian Jahromi, S.~A.~Moosavi, H.~Moradpour, J.~P.~Morais Gra\c{c}a, I.~P.~Lobo, I.~G.~Salako and A.~Jawad,
Phys. Lett. B \textbf{780} (2018), 21-24
doi:10.1016/j.physletb.2018.02.052
[arXiv:1802.07722 [gr-qc]].

\bibitem{Kaniadakis:2005zk}
G.~Kaniadakis,
Phys. Rev. E \textbf{72} (2005), 036108
doi:10.1103/PhysRevE.72.036108
[arXiv:cond-mat/0507311 [cond-mat]].

\bibitem{Drepanou:2021jiv}
N.~Drepanou, A.~Lymperis, E.~N.~Saridakis and K.~Yesmakhanova,
[arXiv:2109.09181 [gr-qc]].

\bibitem{Majhi:2017zao}
A.~Majhi,
Phys. Lett. B \textbf{775} (2017), 32-36
doi:10.1016/j.physletb.2017.10.043
[arXiv:1703.09355 [gr-qc]].

\bibitem{Liu:2021dvj}
Y.~Liu,
doi:10.1209/0295-5075/ac3f52
[arXiv:2112.15077 [gr-qc]].

\bibitem{Witten:1998qj}
E.~Witten,
Adv. Theor. Math. Phys. \textbf{2} (1998), 253-291
doi:10.4310/ATMP.1998.v2.n2.a2
[arXiv:hep-th/9802150 [hep-th]].


\bibitem{Susskind:1998dq}
L.~Susskind and E.~Witten,
[arXiv:hep-th/9805114 [hep-th]].


\bibitem{Fischler:1998st}
W.~Fischler and L.~Susskind,
[arXiv:hep-th/9806039 [hep-th]].


\bibitem{Nojiri:2021iko}
S.~Nojiri, S.~D.~Odintsov and T.~Paul,
Symmetry \textbf{13} (2021) no.6, 928
doi:10.3390/sym13060928
[arXiv:2105.08438 [gr-qc]].

\bibitem{Nojiri:2021jxf}
S.~Nojiri, S.~D.~Odintsov and T.~Paul,
Phys. Lett. B \textbf{825} (2022), 136844
doi:10.1016/j.physletb.2021.136844
[arXiv:2112.10159 [gr-qc]].
 
\bibitem{Li:2004rb}
M.~Li,
Phys.\ Lett.\ B {\bf 603} (2004) 1
doi:10.1016/j.physletb.2004.10.014
[hep-th/0403127].

\bibitem{Li:2011sd}
M.~Li, X.~D.~Li, S.~Wang and Y.~Wang,
Commun. Theor. Phys. \textbf{56} (2011), 525-604
doi:10.1088/0253-6102/56/3/24
[arXiv:1103.5870 [astro-ph.CO]].

\bibitem{Wang:2016och}
S.~Wang, Y.~Wang and M.~Li,
Phys.\ Rept.\  {\bf 696} (2017) 1
doi:10.1016/j.physrep.2017.06.003
[arXiv:1612.00345 [astro-ph.CO]].

\bibitem{Pavon:2005yx}
D.~Pavon and W.~Zimdahl,
Phys.\ Lett.\ B {\bf 628} (2005) 206
doi:10.1016/j.physletb.2005.08.134
[gr-qc/0505020].

\bibitem{Nojiri:2005pu}
S.~Nojiri and S.~D.~Odintsov,
Gen.\ Rel.\ Grav.\  {\bf 38} (2006) 1285
doi:10.1007/s10714-006-0301-6
[hep-th/0506212].

\bibitem{Landim:2022jgr}
R.~G.~Landim,
Phys. Rev. D \textbf{106} (2022) no.4, 043527
doi:10.1103/PhysRevD.106.043527
[arXiv:2206.10205 [astro-ph.CO]].

\bibitem{Zhang:2005yz}
X.~Zhang,
Int.\ J.\ Mod.\ Phys.\ D {\bf 14} (2005) 1597
doi:10.1142/S0218271805007243
[astro-ph/0504586].

\bibitem{Guberina:2005fb}
B.~Guberina, R.~Horvat and H.~Stefancic,
JCAP {\bf 0505} (2005) 001
doi:10.1088/1475-7516/2005/05/001
[astro-ph/0503495].

\bibitem{Elizalde:2005ju}
E.~Elizalde, S.~Nojiri, S.~D.~Odintsov and P.~Wang,
Phys.\ Rev.\ D {\bf 71} (2005) 103504
doi:10.1103/PhysRevD.71.103504
[hep-th/0502082].

\bibitem{Ito:2004qi}
M.~Ito,
Europhys.\ Lett.\  {\bf 71} (2005) 712
doi:10.1209/epl/i2005-10151-x
[hep-th/0405281].

\bibitem{Gong:2004cb}
Y.~g.~Gong, B.~Wang and Y.~Z.~Zhang,
Phys.\ Rev.\ D {\bf 72} (2005) 043510
doi:10.1103/PhysRevD.72.043510
[hep-th/0412218].


\bibitem{Nojiri:2022nmu}
S.~Nojiri, S.~D.~Odintsov and T.~Paul,
Phys. Lett. B \textbf{835} (2022), 137553
doi:10.1016/j.physletb.2022.137553
[arXiv:2211.02822 [gr-qc]].


\bibitem{BouhmadiLopez:2011xi}
M.~Bouhmadi-Lopez, A.~Errahmani and T.~Ouali,
Phys.\ Rev.\ D {\bf 84} (2011) 083508
doi:10.1103/PhysRevD.84.083508
[arXiv:1104.1181 [astro-ph.CO]].

\bibitem{Malekjani:2012bw}
M.~Malekjani,
Astrophys.\ Space Sci.\  {\bf 347} (2013) 405
doi:10.1007/s10509-013-1522-2
[arXiv:1209.5512 [gr-qc]].


\bibitem{Khurshudyan:2016gmb}
M.~Khurshudyan,
Astrophys. Space Sci. \textbf{361} (2016) no.12, 392
doi:10.1007/s10509-016-2981-z

\bibitem{Landim:2015hqa}
R.~C.~G.~Landim,
Int.\ J.\ Mod.\ Phys.\ D {\bf 25} (2016) no.04,  1650050
doi:10.1142/S0218271816500504
[arXiv:1508.07248 [hep-th]].

\bibitem{Gao:2007ep}
C.~Gao, F.~Wu, X.~Chen and Y.~G.~Shen,
Phys.\ Rev.\ D {\bf 79} (2009) 043511
doi:10.1103/PhysRevD.79.043511
[arXiv:0712.1394 [astro-ph]].

\bibitem{Li:2008zq}
M.~Li, C.~Lin and Y.~Wang,
JCAP {\bf 0805} (2008) 023
doi:10.1088/1475-7516/2008/05/023
[arXiv:0801.1407 [astro-ph]].

\bibitem{Anagnostopoulos:2020ctz}
F.~K.~Anagnostopoulos, S.~Basilakos and E.~N.~Saridakis,
[arXiv:2005.10302 [gr-qc]].

\bibitem{Zhang:2005hs}
X.~Zhang and F.~Q.~Wu,
Phys.\ Rev.\ D {\bf 72} (2005) 043524
doi:10.1103/PhysRevD.72.043524
[astro-ph/0506310].

\bibitem{Li:2009bn}
M.~Li, X.~D.~Li, S.~Wang and X.~Zhang,
JCAP {\bf 0906} (2009) 036
doi:10.1088/1475-7516/2009/06/036
[arXiv:0904.0928 [astro-ph.CO]].

\bibitem{Feng:2007wn}
C.~Feng, B.~Wang, Y.~Gong and R.~K.~Su,
JCAP {\bf 0709} (2007) 005
doi:10.1088/1475-7516/2007/09/005
[arXiv:0706.4033 [astro-ph]].

\bibitem{Zhang:2009un}
X.~Zhang,
Phys.\ Rev.\ D {\bf 79} (2009) 103509
doi:10.1103/PhysRevD.79.103509
[arXiv:0901.2262 [astro-ph.CO]].

\bibitem{Lu:2009iv}
J.~Lu, E.~N.~Saridakis, M.~R.~Setare and L.~Xu,
JCAP {\bf 1003} (2010) 031
doi:10.1088/1475-7516/2010/03/031
[arXiv:0912.0923 [astro-ph.CO]].

\bibitem{Micheletti:2009jy}
S.~M.~R.~Micheletti,
JCAP {\bf 1005} (2010) 009
doi:10.1088/1475-7516/2010/05/009
[arXiv:0912.3992 [gr-qc]].


\bibitem{Mukherjee:2017oom}
P.~Mukherjee, A.~Mukherjee, H.~Jassal, A.~Dasgupta and N.~Banerjee,
Eur. Phys. J. Plus \textbf{134} (2019) no.4, 147
doi:10.1140/epjp/i2019-12504-7
[arXiv:1710.02417 [astro-ph.CO]].

\bibitem{Nojiri:2017opc}
S.~Nojiri and S.~Odintsov,
Eur. Phys. J. C \textbf{77} (2017) no.8, 528
doi:10.1140/epjc/s10052-017-5097-x
[arXiv:1703.06372 [hep-th]].




\bibitem{Nojiri:2019skr}
S.~Nojiri, S.~D.~Odintsov and E.~N.~Saridakis,
Eur.\ Phys.\ J.\ C {\bf 79} (2019) no.3,  242
[arXiv:1903.03098 [gr-qc]].

\bibitem{Saridakis:2020zol}
E.~N.~Saridakis,
Phys. Rev. D \textbf{102} (2020) no.12, 123525
doi:10.1103/PhysRevD.102.123525
[arXiv:2005.04115 [gr-qc]].

\bibitem{Barrow:2020kug}
J.~D.~Barrow, S.~Basilakos and E.~N.~Saridakis,
Phys. Lett. B \textbf{815} (2021), 136134
doi:10.1016/j.physletb.2021.136134
[arXiv:2010.00986 [gr-qc]].

\bibitem{Adhikary:2021xym}
P.~Adhikary, S.~Das, S.~Basilakos and E.~N.~Saridakis,
[arXiv:2104.13118 [gr-qc]].

\bibitem{Srivastava:2020cyk}
S.~Srivastava and U.~K.~Sharma,
Int. J. Geom. Meth. Mod. Phys. \textbf{18} (2021) no.01, 2150014
doi:10.1142/S0219887821500146
[arXiv:2010.09439 [physics.gen-ph]].

\bibitem{Bhardwaj:2021chg}
V.~K.~Bhardwaj, A.~Dixit and A.~Pradhan,
New Astron. \textbf{88} (2021), 101623
doi:10.1016/j.newast.2021.101623
[arXiv:2102.09946 [gr-qc]].

\bibitem{Chakraborty:2020jsq}
G.~Chakraborty and S.~Chattopadhyay,
Int. J. Mod. Phys. D \textbf{29} (2020) no.03, 2050024
doi:10.1142/S0218271820500248
[arXiv:2006.07142 [physics.gen-ph]].

\bibitem{Sarkar:2021izd}
A.~Sarkar and S.~Chattopadhyay,
Int. J. Geom. Meth. Mod. Phys. \textbf{18} (2021) no.09, 2150148
doi:10.1142/S0219887821501486

\bibitem{Nojiri:2022aof}
S.~Nojiri, S.~D.~Odintsov and V.~Faraoni,
Phys. Rev. D \textbf{105} (2022) no.4, 044042
doi:10.1103/PhysRevD.105.044042
[arXiv:2201.02424 [gr-qc]].


\bibitem{Nojiri:2022dkr}
S.~Nojiri, S.~D.~Odintsov and T.~Paul,
Phys. Lett. B \textbf{831} (2022), 137189
doi:10.1016/j.physletb.2022.137189
[arXiv:2205.08876 [gr-qc]].



\bibitem{Horvat:2011wr}
R.~Horvat,
Phys. Lett. B \textbf{699} (2011), 174-176
doi:10.1016/j.physletb.2011.04.004
[arXiv:1101.0721 [hep-ph]].

\bibitem{Nojiri:2019kkp}
S.~Nojiri, S.~D.~Odintsov and E.~N.~Saridakis,
Phys. Lett. B \textbf{797} (2019), 134829
doi:10.1016/j.physletb.2019.134829
[arXiv:1904.01345 [gr-qc]].

\bibitem{Paul:2019hys}
T.~Paul,
EPL \textbf{127} (2019) no.2, 20004
doi:10.1209/0295-5075/127/20004
[arXiv:1905.13033 [gr-qc]].

\bibitem{Bargach:2019pst}
A.~Bargach, F.~Bargach, A.~Errahmani and T.~Ouali,
Int. J. Mod. Phys. D \textbf{29} (2020) no.02, 2050010
doi:10.1142/S0218271820500108
[arXiv:1904.06282 [hep-th]].

\bibitem{Elizalde:2019jmh}
E.~Elizalde and A.~Timoshkin,
Eur. Phys. J. C \textbf{79} (2019) no.9, 732
doi:10.1140/epjc/s10052-019-7244-z
[arXiv:1908.08712 [gr-qc]].

\bibitem{Oliveros:2019rnq}
A.~Oliveros and M.~A.~Acero,
EPL \textbf{128} (2019) no.5, 59001
doi:10.1209/0295-5075/128/59001
[arXiv:1911.04482 [gr-qc]].

\bibitem{Mohammadi:2022vru}
A.~Mohammadi,
[arXiv:2203.06643 [gr-qc]].

\bibitem{Chakraborty:2020tge}
G.~Chakraborty and S.~Chattopadhyay,
Int. J. Geom. Meth. Mod. Phys. \textbf{17} (2020) no.5, 2050066
doi:10.1142/S0219887820500668
[arXiv:2006.07143 [physics.gen-ph]].

\bibitem{Nojiri:2020wmh}
S.~Nojiri, S.~D.~Odintsov, V.~K.~Oikonomou and T.~Paul,
Phys. Rev. D \textbf{102} (2020) no.2, 023540
doi:10.1103/PhysRevD.102.023540
[arXiv:2007.06829 [gr-qc]].

\bibitem{Nojiri:2019yzg}
S.~Nojiri, S.~D.~Odintsov and E.~N.~Saridakis,
Nucl. Phys. B \textbf{949} (2019), 114790
doi:10.1016/j.nuclphysb.2019.114790
[arXiv:1908.00389 [gr-qc]].

\bibitem{Brevik:2019mah}
I.~Brevik and A.~Timoshkin,
doi:10.1142/S0219887820500231
[arXiv:1911.09519 [gr-qc]].


\bibitem{Nojiri:2022ljp}
S.~Nojiri, S.~D.~Odintsov and V.~Faraoni,
[arXiv:2208.10235 [gr-qc]].


\bibitem{Padmanabhan:2009vy} 
T.~Padmanabhan,
Rept.\ Prog.\ Phys.\ {\bf 73}, 046901 (2010)
[arXiv:0911.5004 [gr-qc]].

\bibitem{Cai:2005ra}
R.~G.~Cai and S.~P.~Kim,
JHEP \textbf{02} (2005), 050
doi:10.1088/1126-6708/2005/02/050
[arXiv:hep-th/0501055 [hep-th]].

\bibitem{Nojiri:2018ouv}
S.~Nojiri, S.~D.~Odintsov and V.~K.~Oikonomou,
Phys.\ Rev.\ D {\bf 99} (2019) no.4,  044050
doi:10.1103/PhysRevD.99.044050
[arXiv:1811.07790 [gr-qc]].


\bibitem{Odintsov:2022unp}
S.~D.~Odintsov and T.~Paul,
Universe \textbf{8} (2022) no.5, 292
doi:10.3390/universe8050292
[arXiv:2205.09447 [gr-qc]].


\bibitem{Elizalde:2020zcb}
E.~Elizalde, S.~D.~Odintsov, V.~K.~Oikonomou and T.~Paul,
Nucl. Phys. B \textbf{954} (2020), 114984
doi:10.1016/j.nuclphysb.2020.114984
[arXiv:2003.04264 [gr-qc]].


\bibitem{Bamba:2020qdj}
K.~Bamba, E.~Elizalde, S.~D.~Odintsov and T.~Paul,
JCAP \textbf{04} (2021), 009
doi:10.1088/1475-7516/2021/04/009
[arXiv:2012.12742 [gr-qc]].



\bibitem{Nojiri:2022xdo}
S.~Nojiri, S.~D.~Odintsov and T.~Paul,
Phys. Dark Univ. \textbf{35} (2022), 100984
doi:10.1016/j.dark.2022.100984
[arXiv:2202.02695 [gr-qc]].


\bibitem{Odintsov:2020sqy}
S.~D.~Odintsov, V.~K.~Oikonomou and F.~P.~Fronimos,
Nucl. Phys. B \textbf{958} (2020), 115135
doi:10.1016/j.nuclphysb.2020.115135
[arXiv:2003.13724 [gr-qc]].


\bibitem{Hwang:2005hb}
J.~c.~Hwang and H.~Noh,
Phys.\ Rev.\ D {\bf 71} (2005) 063536
doi:10.1103/PhysRevD.71.063536
[gr-qc/0412126].

\bibitem{Noh:2001ia}
H.~Noh and J.~c.~Hwang,
Phys.\ Lett.\ B {\bf 515} (2001) 231
doi:10.1016/S0370-2693(01)00875-9
[astro-ph/0107069].


\bibitem{Akrami:2018odb}
Y.~Akrami {\it et al.} [Planck Collaboration],
arXiv:1807.06211 [astro-ph.CO].


\end{thebibliography}
\end{document}